\def\fdg{\hbox{$.\!\!^\circ$}}
\def\fmn{\hbox{$.\!^{'}$}}
\documentclass{aa}
\usepackage{epsfig}
\usepackage {graphicx}
\usepackage{amssymb}
\usepackage{amsmath}
\usepackage{multirow}
\usepackage {txfonts}
\usepackage{natbib}

\begin{document}


\title{Interstellar gas within $\sim 10$~pc of Sgr~A$^*$}
\author{Katia Ferri\`ere\inst{1}}
\offprints{Katia Ferri\`ere}
\institute{$^{1}$ IRAP, Universit\'e de Toulouse, CNRS,
14 avenue Edouard Belin, F-31400 Toulouse, France }
\date{Received  ; accepted }
\titlerunning{Interstellar gas near Sgr~A$^*$}
\authorrunning{Katia Ferri\`ere}

\abstract
{}
{We seek to obtain a coherent and realistic three-dimensional picture 
of the interstellar gas out to about 10~pc of the dynamical center 
of our Galaxy, which is supposed to be at Sgr~A$^*$.}
{We review the existing observational studies on the different
gaseous components that have been identified near Sgr~A$^*$,
and retain all the information relating to
their spatial configuration and/or physical state.
Based on the collected information, we propose a three-dimensional 
representation of the interstellar gas, which describes each component 
in terms of both its precise location and morphology 
and its thermodynamic properties.}
{The interstellar gas near Sgr~A$^*$ can represented 
by five basic components, which are, by order of increasing size: 
(1) a central cavity with roughly equal amounts of warm ionized 
and atomic gases,
(2) a ring of mainly molecular gas,
(3) a supernova remnant filled with hot ionized gas,
(4) a radio halo of warm ionized gas and relativistic particles,
and (5) a belt of massive molecular clouds.
While the halo gas fills $\approx 80\%$ of the studied volume,
the molecular components enclose $\approx 98\%$ of the interstellar mass.
}
{}

\keywords{ISM: structure - Galaxy: center - Galaxy: nucleus - 
ISM: general - ISM: kinematics and dynamics - ISM: supernova remnants}

\maketitle

\section{\label{intro}Introduction}

There is now compelling evidence, largely based on measured stellar orbits, 
that a massive black hole with mass $\approx 4 \times 10^6~M_\odot$
resides at the dynamical center of our Galaxy, which is traditionally 
identified with the compact nonthermal radio source Sagittarius A$^*$ 
(Sgr~A$^*$) \citep[see][for a recent, comprehensive review]{genzel&eg_10}.
Sgr~A$^*$ sits in the heart of a dense cluster of young, 
massive and luminous stars, which are concentrated within 
the central parsec and are distributed in two relatively thick disks 
that are highly inclined toward each other and rotate in opposite directions
\citep{krabbe&gdr_91, krabbe&gen_95, paumard&gmn_06}.
This central star cluster, which has a few bright condensations
(notably, the IRS~16 complex at the very center), 
is the source of intense UV radiation and powerful stellar winds.
Also present around Sgr~A$^*$, but extending much farther out,
is a cluster of old and evolved, cool stars,
with nearly isotropic distribution and slow, solid-body rotation
\citep{blum&rso_03, trippe&ggb_08, schodel&me_09}.
The old cluster dominates the stellar mass by far, 
with $\sim 10^6~M_\odot$ inside the central 1~pc \citep{schodel&me_09}, 
as opposed to $\lesssim 1.5 \times 10^4~M_\odot$ in the young cluster 
\citep{paumard&gmn_06}.
Like in the rest of the Galaxy, all stars are immersed in 
an interstellar medium (ISM), which is essentially made of gas 
(in molecular, atomic and ionized forms) and dust.

The few-parsec region surrounding Sgr~A$^*$ is of indisputable interest,
not only in its own right, because it constitutes a unique, extremely 
complex and highly interacting Galactic environment,
but also from a broader perspective, 
because it represents (by far) the nearest and, therefore, 
most easily accessible example of a galactic nucleus, and as such 
may be the key to understanding the energetic processes taking place
in galactic nuclei in general.
This is why the central few parsecs have long been the target 
of numerous observations over a wide range of frequencies.
Recent years have witnessed dramatic progress 
at both the low-energy (radio and infrared) and high-energy 
(X-ray and $\gamma$-ray) ends of the electromagnetic spectrum.
Yet, despite the wealth of observational data that have now become 
available, achieving a clear and complete three-dimensional view 
of the innermost Galactic region remains a challenging task,
due in large part to the difficulty in positioning the observed 
features along the line of sight.

In this context, we will try to unravel at best the intricate spatial
distribution of the interstellar gas within $\sim 10$~pc of Sgr~A$^*$, 
and we will describe the emerging picture by way of a simplified
three-dimensional gas representation, which we mean to be as realistic 
as possible.
Let us specify from the outset that we will truly focus 
on the interstellar gas.
Stars will only be alluded to for their direct impact 
on the interstellar gas, and interstellar magnetic fields 
will be tackled in a separate paper.
Let us also emphasize that our purpose is not to provide 
a comprehensive overview of the interstellar gas in the region of interest.
Instead, we will present what we feel are the most important 
and directly relevant observational studies.
We will discuss their sometimes divergent results and extract 
the useful pieces of information that can serve as building blocks
for our gas representation. 
We will then assemble all the pieces of information 
and try to reconstruct the overall puzzle.
The existing observations will be presented in Section~\ref{obs_gas},
while our gas representation will be the subject of Section~\ref{model_gas}.

Owing to the considerable uncertainties in the observational results and 
in their interpretations, which are reflected in the disparate conclusions 
reached by different authors, 
our gas representation will necessarily be approximate.
However, we hope that it can be used as an observational input 
to theoretical studies that deal with Sgr~A$^*$ and its surroundings.
One such study that we have undertaken in parallel to the present work
concerns the propagation and annihilation of positrons from Sgr A$^*$ 
(Jean et al., in preparation) -- an important investigation in direct need
of a realistic and reasonably accurate description of the interstellar gas.

\begin{figure}[!htb]
\centering
\hspace*{-0.8cm}
\includegraphics[scale=0.65]{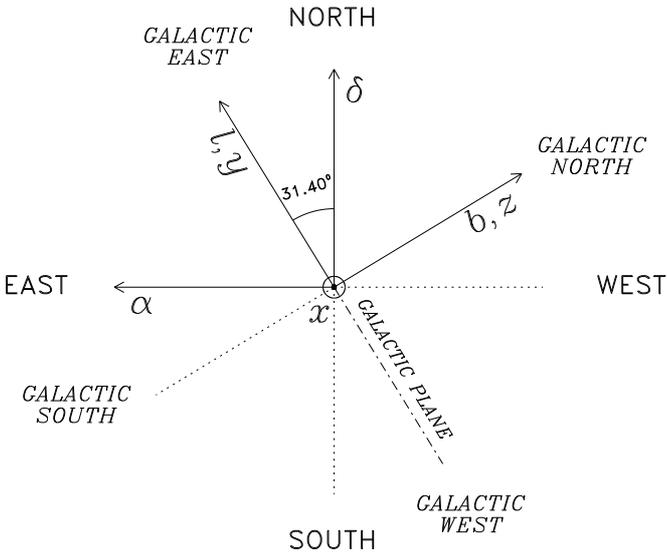}
\caption{
\label{axes}
Different coordinate systems used in this paper, 
shown in the plane of the sky:
$\alpha$ (right ascension) and $\delta$ (declination) are
the equatorial coordinates;
$l$ (longitude) and $b$ (latitude) are the Galactic coordinates;
and $x$ (toward the Sun), $y$ and $z$ are the Galactocentric 
cartesian coordinates.
See text for more details.
}
\end{figure}

Unless stated otherwise, the observational maps presented here
will be discussed in the equatorial coordinate system defined by
right ascension, $\alpha$, and declination, $\delta$.
In this system, east/west refers to the direction 
of increasing/decreasing $\alpha$ and north/south 
to the direction of increasing/decreasing $\delta$
(see Figure~\ref{axes}).
We will also use the Galactic coordinate system defined by 
longitude, $l$, and latitude, $b$,
where Galactic east/west refers to the direction 
of increasing/decreasing $l$ and Galactic north/south 
to the direction of increasing/decreasing $b$.
At the position of Sgr~A$^*$, the trace of the Galactic plane 
($b = 0^\circ$) in the plane of the sky is at position angle 
$31\fdg40$ east of north \citep[in J2000;][]{reid&b_04},
so that there is a $58\fdg60$ angle between the $(\alpha,\delta)$ 
and $(l,b)$ systems.

Our gas representation will be described in terms of the Galactocentric 
cartesian coordinates, $(x,y,z)$,
where $x$ is the horizontal (i.e., parallel to the Galactic plane) 
coordinate along the line of sight to the Sun (positive toward the Sun),
$y$ the horizontal coordinate in the plane of the sky
(positive toward Galactic east)
and $z$ the vertical (i.e., perpendicular to the Galactic plane)
coordinate (positive toward Galactic north).
For consistency with our previous papers on the Galactic center (GC) region, 
we will adopt $r_\odot = 8.5$~kpc for the Galactocentric radius of the Sun,
such that angular separations of $1'$ and $1''$ translate into 
linear separations near Sgr~A$^*$ of $2.5$~pc and $0.04$~pc, respectively.

The coordinates of Sgr~A$^*$ in the three different systems are
$(\alpha_{\rm A^*}, \delta_{\rm A^*}) =
(17^{\rm h} 45^{\rm m} 40.\!^{\rm s}04, -29^\circ 00' 28.\!''\!1)$ 
\citep[in J2000;][]{reid&b_04},
$(l_{\rm A^*}, b_{\rm A^*}) = 
(-0^\circ 03' 20.\!''\!5, -0^\circ 02' 46.\!''\!3)$
\citep[as calculated using the $(\alpha, \delta)$ coordinates 
of the origin of the $(l, b)$ system given by][]{reid&b_04}
and $(x_{\rm A^*}, y_{\rm A^*}, z_{\rm A^*}) = 0$ (by construction).
In the following, angular offsets with respect to Sgr~A$^*$ will be denoted 
by $(\Delta \alpha, \Delta \delta)$ in the equatorial system
and by $(\Delta l, \Delta b)$ in the Galactic system.

\section{\label{obs_gas}Observations of the interstellar gas}

The interstellar gas in the immediate vicinity of Sgr~A$^*$
has a complicated morphology \citep[see, e.g.,][for early reviews]
{morris&s_96, mezger&dz_96}. 
In brief, Sgr~A$^*$ is embedded in a $\approx (2 - 3)$~pc sized cavity,
the "Central Cavity", which was originally identified as a filamentary 
H\,{\sc ii} region named Sgr~A West.
This cavity has most likely been evacuated by stellar winds 
and photo-ionized by UV radiation from the central star cluster.
Encircling the Central Cavity is an asymmetric torus of neutral
(mainly molecular) gas and dust, usually referred to as 
the Circumnuclear Disk (CND) or, more appropriately, 
the Circumnuclear Ring (CNR).
The CNR is generally interpreted as being part of an accretion disk around 
the central massive black hole, even though its pronounced asymmetry
suggests that it is a transient feature.
Both the Central Cavity and the CNR lie, in projection (onto the plane
of the sky), inside a $\approx 10$~pc scale nonthermal radio shell 
called Sgr~A East and widely believed to be a supernova remnant (SNR).
Sgr~A East, in turn, is surrounded by a $\approx 20$~pc diameter 
radio halo, which is probably composed of a mixture of
warm ionized gas (thermal component)
and relativistic particles (nonthermal component).
Finally, a belt of massive molecular clouds around Sgr~A East
stretches over $\approx 30$~pc along the Galactic plane; 
most prominent amongst these clouds are the well-known M$-$0.02$-$0.07 
(or $50~{\rm km~s}^{-1}$) and M$-$0.13$-$0.08 (or $20~{\rm km~s}^{-1}$) 
giant molecular clouds (GMCs) located east and south, respectively, 
of Sgr~A East.
We now proceed to describe each of the above structural components 
in more detail.

\subsection{\label{obs_cavity}The Central Cavity}

\begin{figure}[!t]
\centering
\includegraphics[width=0.48\textwidth]{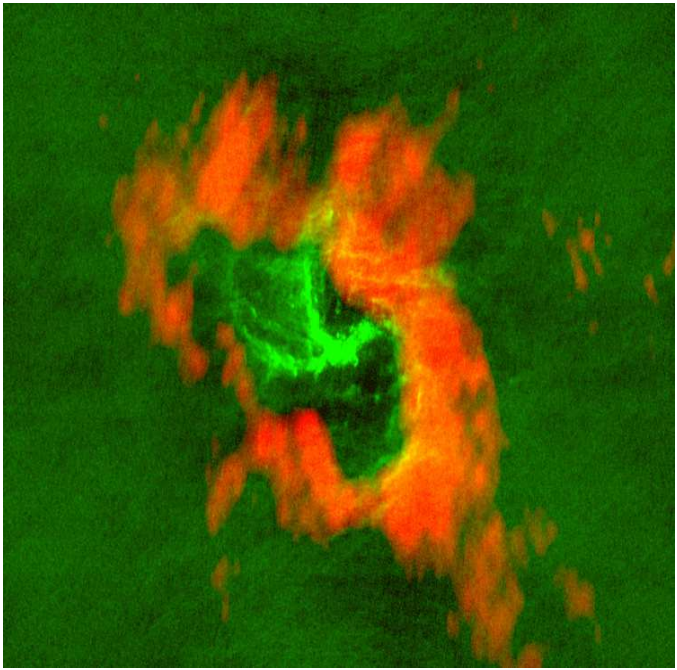}
\caption{
\label{CNR}
Composite image showing (in green) the 3.6~cm radio continuum emission
from warm ionized gas in the Sgr~A West H\,{\sc ii} region,
with the three-arm Minispiral emerging very clearly,
and (in red) the 3.4~mm HCN $J = 1 \to 0$ line emission
from the surrounding Circumnuclear Ring (CNR).
The radio continuum data are from \cite{yusef&bwr_08}
and the HCN data from \cite{wright&cmh_01}.
Figure credit: Farhad Yusef-Zadeh.
}
\end{figure}

The Sgr~A West H\,{\sc ii} region appears in projection 
as a three-arm spiral, commonly known as the Minispiral 
and composed of the so-called Northern Arm, Eastern Arm and Western Arc 
as well as a short east-west Bar that connects the southern end 
of the Northern Arm and the western end of the Eastern Arm 
to the Western Arc (see Figure~\ref{CNR}).
It is now generally accepted that the Western Arc is the photo-ionized 
inner edge of the western part of the CNR
\citep[e.g.,][]{lo&c_83, serabyn&l_85, gusten&gwj_87, roberts&g_93},
while the Northern and Eastern Arms are the photo-ionized surfaces
of tidally stretched streamers of material falling in toward 
the central massive black hole
\citep[e.g.,][]{lo&c_83, serabyn&l_85, ekers&gsg_83, davidson&wwl_92, 
jackson&ggr_93}.
For completeness, we should mention that 
the Eastern Arm has also been suggested, by analogy with the Western Arc, 
to be the photo-ionized inner edge of the eastern part of the CNR
\citep[e.g.,][]{aitken&smr_98, shukla&ys_04}.
However, this suggestion appears to be incompatible with the finding
that the Eastern Arm is nearly perpendicular to the CNR
\citep{latvakoski&sgh_99} and to the Western Arc \citep{zhao&bmd_10}.

\cite{lo&c_83} presented a VLA 6~cm radio continuum map of the central 3~pc 
of the Galaxy, which clearly shows the spiral structure of Sgr~A West.
They estimated the total mass of ionized gas in the Minispiral
at $\approx 60~M_\odot$, 
with $\approx 10~M_\odot$ in the Northern Arm,
$\approx 10~M_\odot$ in the Eastern Arm,
$\approx 35~M_\odot$ in the Western Arc 
(which they referred to as the south arm)
and $\approx 5~M_\odot$ in the Bar (which they referred to as the west arm).
They also estimated the electron density at 
$\approx 5 \times 10^4~{\rm cm}^{-3}$ in the brightest clumps
and $\approx 10^3~{\rm cm}^{-3}$ in the lower-brightness features.
\cite{ekers&gsg_83}, who mapped the Sgr~A West region with the VLA
at 2~cm, 6~cm and 20~cm, found that the Minispiral has projected dimensions 
$\approx 60'' \times 25''$ ($2.5~{\rm pc} \times 1.0~{\rm pc}$),
with the long dimension along the Galactic plane,
and for each of the three spiral arms, they derived 
an average emission measure $\approx 4 \times 10^6~{\rm pc~cm}^{-6}$,
an electron density $\approx 4 \times 10^3~{\rm cm}^{-3}$,
and a total mass of ionized gas $\approx 15~M_\odot$
(assuming a filling factor of unity).
They also detected diffuse emission from Sgr~A West,
over an area $\approx 75'' \times 40''$ ($3.1~{\rm pc} \times 1.6~{\rm pc}$),
but they argued that this diffuse emission is mainly nonthermal
and that most of the thermal emission from Sgr~A West (at least at 6~cm)
really comes from the Minispiral.
Note that both \cite{lo&c_83} and \cite{ekers&gsg_83} assumed 
$r_\odot = 10$~kpc, so that their estimated masses and densities 
need to be rescaled to our adopted value of $r_\odot = 8.5$~kpc.\footnote{
As a general rule, distances scale as $r_\odot$, surface densities 
as $r_\odot^0$, volume densities as $r_\odot^{-1}$ and masses 
as $r_\odot^2$.
However, when volume densities and masses are inferred from 
emission measures, they scale instead as $r_\odot^{-0.5}$ and 
$r_\odot^{2.5}$, respectively.
}

Similarly, by analyzing the VLA 2~cm radio continuum map 
of \cite{brown&l_84}, \cite{beckert&dmz_96} found that 
the Minispiral is superimposed onto an extended Gaussian component
with projected {\it FWHM} dimensions $\approx 70'' \times 52''$
($2.9~{\rm pc} \times 2.1~{\rm pc}$).
However, they attributed the whole extended component to thermal
(free-free) emission, which, for an electron temperature $\approx 6\,000$~K,
gave them a central emission measure $\approx 2 \times 10^6~{\rm pc~cm}^{-6}$.
Assuming, in addition, a line-of-sight depth $\approx 1$~pc 
and a filling factor of unity, they derived 
an electron density $\approx 1.4 \times 10^3~{\rm cm}^{-3}$ 
and a total H$^+$ mass $\approx 250~M_\odot$ for the extended component.
For the Minispiral, they assumed a line-of-sight depth $\approx 0.1$~pc
and obtained an electron density $\approx 10^4~{\rm cm}^{-3}$
and a total H$^+$ mass $\approx 9~M_\odot$.
Hence, \cite{beckert&dmz_96} came to the conclusion that, 
despite its spiral appearance, Sgr~A West has most of its mass 
spread out throughout its volume.
They also remarked that their derived emission measure of the extended 
component is perfectly consistent with the emission measure 
$\approx 10^6~{\rm pc~cm}^{-6}$ needed to explain the low-frequency
turnover in the Sgr~A$^*$ radio spectrum by foreground free-free absorption,
if Sgr~A$^*$ lies in the middle of Sgr~A West and in front of the Minispiral.

\cite{yusef&me_89} made higher-resolution VLA 2~cm and 6~cm radio 
continuum observations of Sgr~A West, which brought out a number 
of fine-scale morphological details.
For instance, they detected the stellar wind from the red supergiant IRS~7, 
located $\approx 0.25$~pc north of Sgr~A$^*$,
and they suggested that this wind is photo-ionized externally
by the UV radiation bathing the Central Cavity.
They also observed the so-called Minicavity
\citep[first described by][]{morris&y_87}
-- a nearly circular, $\approx 0.08$~pc diameter hole
in the distribution of ionized gas in the east-west Bar,
centered $\approx 0.14$~pc southwest of Sgr~A$^*$.
They naturally proposed that the Minicavity was swept out by a spherical 
stellar wind, although the most obvious stellar candidate, the IRS~16 complex, 
lies at the northeastern periphery of the Minicavity, not at its center.
Other possible scenarios were later put forward.
\cite{roberts&yg_96} suggested that the Minicavity is the ionized
component of a compact molecular cloud moving along an orbit 
that passes very close to Sgr~A$^*$.
Along different lines, \cite{lutz&kg_93} invoked a fast wind 
from one or more nearby sources which would blow into the Bar streamer 
and produce an expanding gas bubble there. 
\cite{melia&cy_96} proposed a more elaborate model, in which 
the central massive black hole gravitationally focuses the wind 
from IRS~16, partially accretes from it, and expels the rest
in a collimated flow whose collision with the Bar streamer
creates the Minicavity.

Sgr~A West was also observed in various emission lines, 
starting with infrared lines such as 
the $12.8~\mu$m [Ne\,{\sc ii}] fine-structure line 
\citep{lacy&btg_79, lacy&tgh_80, serabyn&l_85, serabyn&ltb_88, lacy&as_91},
the $2.06~\mu$m He\,{\sc i} line 
\citep{hall&ks_82, geballe&kbw_91, paumard&mm_04},
the $2.17~\mu$m H\,{\sc i} Br$\gamma$ recombination line
\citep{wright&mb_89, geballe&kbw_91, herbst&bfp_93, paumard&mm_04}, etc.
The early infrared line emission images
\citep[see, e.g.,][]{wright&mb_89, lacy&as_91} were found to
closely resemble the 6~cm radio continuum map of \cite{lo&c_83}.
Besides, \cite{lacy&as_91} were able to reproduce the morphology
and kinematics of much of the [Ne\,{\sc ii}]-emitting ionized gas 
with a one-armed linear spiral containing the Western Arc and 
the Northern Arm (but excluding the Eastern Arm and most of the Bar),
along which the gas is in nearly circular Keplerian rotation about
Sgr~A$^*$.
The [Ne\,{\sc ii}] data of \cite{lacy&as_91} were subsequently
re-examined by \cite{vollmer&d_00}, who, in addition to circular rotation,
allowed for turbulent motions plus slow radial accretion.
They found that the Western Arc, the Northern Arm and part of the Bar
are located in a single plane (the plane of the CNR),
whereas the Eastern Arm is actually composed of two distinct pieces 
belonging to two different planes, one of which also encloses 
the rest of the Bar.
Aside from the dense gas confined to the Minispiral, tenuous gas
appears to fill the plane of the Western Arc + Northern Arm
and possibly one of the planes of the Eastern Arm.

In parallel to infrared emission lines, which are plagued by
interstellar dust extinction, radio emission lines have proven
particularly valuable since the early work of
\cite{vangorkom&sb_85} and \cite{schwarz&bv_89}.
\cite{roberts&g_93} carried out VLA observations of the 8.3~GHz (3.6~cm)
H92$\alpha$ recombination line to investigate the global kinematics 
and temperature distribution of Sgr~A West.
They found that the Western Arc, the Northern Arm and the extended bar
(composed of the Eastern Arm, the Bar and its linear extension
to the northwest) constitute three distinct kinematic entities.
In contrast to the Western Arc, which appears to be in near circular rotation
about Sgr~A$^*$, the Northern Arm and the extended bar do not appear to
have significant circular motions.
\cite{roberts&g_93} also derived the electron temperature in Sgr~A West,
by combining their H92$\alpha$ line data with a 8.3~GHz continuum map
and making the assumptions (all found to be satisfied in the case at hand)
that the emitting gas is in local thermodynamic equilibrium (LTE), 
the continuum emission is thermal free-free,
the continuum and recombination-line emissions are optically thin,
and pressure broadening is negligible.
Under these conditions, the line-to-continuum ratio implies
an electron temperature $\approx 7\,000$~K that is
approximately uniform throughout Sgr~A West.

Because radio recombination lines at centimeter wavelengths can be 
dominated by stimulated emission and affected by pressure broadening, 
\cite{shukla&ys_04} observed the Sgr~A West region in the 92~GHz (3.3~mm) 
H41$\alpha$ recombination line.
At this higher frequency, the continuum and recombination-line emissions
are believed to arise mostly in denser ionized gas.
The morphology of the ionized gas in the 92~GHz continuum and
H41$\alpha$ line images is very similar to that in the 8.3~GHz 
continuum image of \cite{roberts&g_93}.
The Northern Arm appears to be $\approx 0.9$~pc long, 
the Eastern Arm $\approx 0.4$~pc, 
the Western Arc $\approx 1.5$~pc,
the Bar $\approx 1.1$~pc, 
and the four spiral features are roughly 0.1~pc wide.
Under the same assumptions as those made by \cite{roberts&g_93},
\cite{shukla&ys_04} also derived a fairly uniform 
electron temperature $\approx 7\,000$~K. 
At this temperature, the intensity of the continuum 
(supposedly thermal free-free) emission from the arms
corresponds to an emission measure $\approx 2.5 \times 10^7~{\rm pc~cm}^{-6}$, 
which, if the line-of-sight thickness of the arms is comparable to their width
($\approx 0.1$~pc) and if the ionized gas in the arms is smoothly distributed
(filling factor $\approx 1$), implies an electron density 
$\approx 1.6 \times 10^4~{\rm cm}^{-3}$ in the arms.
The resulting H$^+$ masses are
$\approx 2.8~M_\odot$ in the Northern Arm,
$\approx 1.2~M_\odot$ in the Eastern Arm,
$\approx 4.6~M_\odot$ in the Western Arc,
$\approx 3.4~M_\odot$ in the Bar,
and hence $\approx 12~M_\odot$ in the entire Minispiral.\footnote{
Strictly speaking, these estimates suppose that the Minispiral lies 
in the plane of the sky, which is almost certainly incorrect. 
In reality, if a section of an arm makes an angle $\theta$ 
with the plane of the sky, its estimated electron density and H$^+$ mass
(at given emission measure) should be multiplied by 
$(\cos \theta)^{1/2}$ and $(\cos \theta)^{-1/2}$, respectively. 
However, a more important source of error is the uncertainty in the true
width/thickness of the arms, which could possibly be as much as
2 times larger than quoted by \cite{shukla&ys_04}.
In this case, the electron density would be lower by a factor
$2^{0.5} \simeq 1.4$ and the H$^+$ masses larger by a factor
$2^{1.5} \simeq 2.8$.
}
Otherwise, the kinematics of the H41$\alpha$-emitting gas
are essentially the same as those of the H92$\alpha$-emitting gas
\citep[see][]{roberts&g_93}.

A more complex picture of Sgr~A West emerged from the work
of \cite{paumard&mm_04}, who observed the inner parts of the Minispiral
in the $2.17~\mu$m H\,{\sc i} Br$\gamma$ and $2.06~\mu$m He\,{\sc i} 
emission lines.
A kinematic analysis of their data led them to identify
nine coherent velocity components,
comprising both extended, continuous flows and smaller, isolated patches.
The most prominent component is the Northern Arm, which spreads well beyond 
the namesake filament seen in intensity maps,
forming a wedge extending all the way over to the Eastern Arm.
The latter is divided into three parts: two roughly parallel
elongated features, named the Ribbon and the Eastern Bridge, 
and a smaller patchy feature at the western end, named the Tip.
The Bar is straight and extends from the Ribbon to the Western Arc,
which is only partly visible in the present, limited field of view.
Then come three small patchy features, named the Western Bridge, 
the Northern Arm Chunk and the Bar Overlay.
Focusing on the Northern Arm, \cite{paumard&mm_04} showed that 
its kinematics could be mostly modeled with a system of Keplerian orbits 
about Sgr~A$^*$;
these orbits are all close to the plane of the CNR, albeit not perfectly
coplanar, such that they form a three-dimensional, saddle-shaped surface 
-- like the inner surface of a torus.
This kind of geometry would naturally come about if an infalling cloud
was tidally stretched by the central massive black hole 
and had its inward-facing side photo-ionized by hot stars 
from, e.g., the IRS~16 complex.
The warping of the flow surface would then give rise to orbit crowding 
in the plane of the sky, precisely along the bright filament seen 
in intensity maps.
Finally, from the detection of two spots of extinction in the flux maps
of the Northern Arm and the Bar, \cite{paumard&mm_04} concluded that,
along the line of sight, the Bar lies behind the Northern Arm, 
which itself lies behind the Eastern Bridge.

Complementing the radial velocities extracted from infrared and radio 
spectral lines are the transverse velocities associated with proper motions.
\cite{yusef&rb_98} measured proper motions of ionized gas in Sgr~A West,
based on VLA 2~cm radio continuum observations made
at three epochs over a nine-year period.
Their measurements revealed the existence of several features 
(notably, a head-tail structure dubbed the "bullet") 
with transverse velocities greater than the local escape velocities.
The authors combined their measured transverse velocities 
with existing radial velocities from H92$\alpha$ line data,
whereby they found that most, and probably almost all, 
of the total velocities exceed the local escape velocities.
From this, they concluded that ionized gas in the Northern Arm
is probably on an unbound orbit -- in agreement with \cite{roberts&yg_96}, 
who attributed their measured H92$\alpha$ radial velocities in the Minicavity 
to ionized gas being on a hyperbolic orbit about Sgr~A$^*$.
\citeauthor{yusef&rb_98}'s (\citeyear{yusef&rb_98})conclusion, 
which contradicts the widely accepted view that the Northern Arm 
is a tidally stretched streamer of infalling gas,
could be explained if an energetic phenomenon, a few parsecs away
from Sgr~A$^*$, pushed a gas cloud into the Galactic center.

Similarly, \cite{muzik&esm_07} performed the first proper motion
measurements of infrared dust filaments in the Minispiral
and showed that their shapes and velocities are {\it not} consistent
with pure Keplerian rotation about Sgr~A$^*$.
Instead, they could result from the dynamical interaction
between a fast GC outflow and the Minispiral, 
where the GC outflow could either originate in the central cluster 
of young mass-losing stars, or emanate from the accreting black hole, 
possibly in the form of collimated jets \citep[e.g.,][]{yuan_06},
or even arise from a combination of both.
Thus \citeauthor{muzik&esm_07}'s (\citeyear{muzik&esm_07}) filaments
provide a new piece of circumstantial evidence for the existence 
of a GC outflow, adding to the Minicavity, which seems to be connected 
to Sgr~A$^*$ by a chain of plasma blobs \citep{wardle&y_92, melia&cy_96};
the bow-shock structure of the extended ionized envelope of IRS~7, 
with the apex facing toward Sgr~A$^*$ or IRS~16 \citep{yusef&m_92}, 
and the associated cometary tail of ionized gas pointing directly 
away from Sgr~A$^*$ \citep{yusef&m_91};
the similar bow-shock/cometary-tail morphology of IRS~3 
\citep{viehmann&esm_05};
the observed waviness of the Northern Arm \citep{yusef&w_93};
the narrow channel of low interstellar extinction running
northeast-southwest through Sgr A*, with aligned cometary features
\citep{schodel&eam_07}; etc.

The dynamics of the three ionized streams in Sgr~A West
were further studied by \cite{zhao&mga_09}, who combined 
proper motion measurements for 71 compact H\,{\sc ii} features 
with radial velocity measurements from archival H92$\alpha$ line data.
They were able to model the three ionized streams with three bundles 
of Keplerian elliptical orbits about Sgr~A$^*$, all three bundles being 
confined within the central 3~pc.
They confirmed that the Western Arc stream is nearly circular,
while the Northern and Eastern Arm streams have high eccentricities,
and they suggested that the latter may collide in the Bar region,
which they located mostly behind Sgr~A$^*$.
They also found some support for \citeauthor{liszt03}'s 
(\citeyear{liszt03}) suggestion
that the Eastern Arm and the Bar together form a single streamer.
For future reference, the modeled orbital parameters of the three 
ionized streams (rescaled to $r_\odot = 8.5$~kpc and adjusted to 
our line-of-sight vector pointing toward the Sun) are as follows:
the Northern Arm has semimajor axis $a = 1.05$~pc, 
eccentricity $e = 0.83$, inclination (between the angular momentum 
vector and the line-of-sight vector) $i = 41^\circ$
and total length (calculated from the quoted range of true anomaly)
$l = 2.7$~pc;
the Eastern Arm (or Eastern Arm + Bar) has $a = 1.5$~pc, $e = 0.82$, 
$i = 58^\circ$ and $l = 3.5$~pc;
and the Western Arc has $a = 1.2$~pc, $e = 0.20$, $i = 63^\circ$
and $l = 3.5$~pc.

Relying on the Keplerian model of \cite{zhao&mga_09},
\cite{zhao&bmd_10} drew a three-dimensional view
of the three ionized streams, which clearly shows that
the Northern Arm and Western Arc are nearly coplanar,
that their mean orbital plane is nearly perpendicular 
to the orbital plane of the Eastern Arm,
and that the Eastern Arm collides with the Northern Arm 
in the Bar region, just behind Sgr~A$^*$.
More importantly, \cite{zhao&bmd_10} presented new observations 
of the 232~GHz (1.3~mm) H30$\alpha$ recombination line,
which they interpreted together with previous H92$\alpha$ line 
and 22~GHz continuum measurements in the framework of an isothermal,
homogeneous, non-LTE H\,{\sc ii} model,
to determine the physical parameters of the high-density ionized gas 
at selected positions (toward known infrared sources)
along the Northern and Eastern Arms.
They obtained electron kinetic temperatures in the range 
$\approx (5\,000 - 11\,000)$~K and electron densities in the range
$\approx (10^4 - 10^5)~{\rm cm}^{-3}$, with values up to 
$\approx 13\,000$~K and $\approx 2 \times 10^5~{\rm cm}^{-3}$
in the Bar region.
The higher electron temperatures and densities in the Bar region
could result either from gas heating and compression by powerful winds 
from stellar clusters such as IRS~16 and IRS~13
or from the collision between the Northern and Eastern Arms.

\smallskip
\centerline{*****}

In addition to warm ionized gas, the Central Cavity 
also contains substantial amounts of neutral atomic gas, 
detectable through atomic line emission 
\citep{genzel&ctw_85, poglitsch&sgh_91, jackson&ggr_93} 
and through dust thermal continuum emission 
\citep{davidson&wwl_92, zylka&mwd_95, telesco&dw_96, latvakoski&sgh_99}.
On the other hand, little molecular gas seems to be present,
except possibly for a hot and dense molecular component detected 
in NH$_3$ $(6,6)$ emission very near Sgr~A$^*$ \citep{herrnstein&h_02}.
The general lack of molecular gas can be understood if any initially 
molecular cloud inside the Central Cavity has been largely photo-dissociated 
by the intense ambient UV radiation 
\citep{jackson&ggr_93, latvakoski&sgh_99, shukla&ys_04}.

The $63~\mu$m [O\,{\sc i}] $^3P_1 \to \ ^3P_2$ line observations
of \cite{jackson&ggr_93} indicate that neutral atomic gas 
exhibits two prominent emission peaks on opposite sides of Sgr~A$^*$.
The northern peak is part of a north-south, $\approx 1'$ (2.5~pc) 
long ridge of [O\,{\sc i}] emission, which the authors interpreted 
as a streamer of neutral atomic gas falling through a gap in the CNR 
into the Central Cavity, between the Minispiral's Northern and Eastern Arms.
The latter, they suggested, could simply be bright rims of ionized gas
at the surface of this neutral "Northern Streamer"
-- a suggestion already made by \cite{davidson&wwl_92}, who observed 
the Northern Streamer in dust far-infrared emission.
Alternatively, only the Northern Arm would border the Northern Streamer, 
while the Eastern Arm would be an ionized rim at the surface 
of another infalling neutral streamer.\footnote{
Clearly, the recent work of \cite{zhao&bmd_10}, 
which finds the Northern and Eastern Arms to be on nearly perpendicular 
orbits, makes the second scenario much more likely.
}
The southern [O\,{\sc i}] emission peak, for its part, is close to
(but apparently slightly outside) the most prominent radio continuum
emission peak in the Western Arc, which, we recall, is generally believed 
to be the ionized inner edge of the western portion of the CNR.
By using their $63~\mu$m [O\,{\sc i}] data toward the northern peak 
together with previous $146~\mu$m [O\,{\sc i}] and $158~\mu$m [C\,{\sc ii}] 
data \citep[from][]{genzel&ctw_85, poglitsch&sgh_91} as input to model 
calculations of collisional excitation and radiative transport, 
\cite{jackson&ggr_93} estimated that neutral (presumably atomic) gas 
inside the Central Cavity has a temperature $\approx (170 \pm 70)$~K,
a true hydrogen density $\approx 3 \times 10^5~{\rm cm}^{-3}$,
a space-averaged hydrogen density $\approx 1.6 \times 10^3~{\rm cm}^{-3}$
(beam-averaged hydrogen column density in a $55''$ beam divided by 
an assumed line-of-sight pathlength of 1~pc; note, however, 
that the value reported in their Table~2 is too low by a factor of 2)
and a total hydrogen mass (associated with the Northern and Eastern Arms)
$\approx 300~M_\odot$.

\cite{latvakoski&sgh_99} found that the three arms of the Minispiral 
seen in thermal radio continuum emission have counterparts
in dust far-infrared continuum emission, which tend to lie 
$\approx 1'' - 3''$ farther from Sgr~A$^*$.
This configuration, they explained, is consistent with the radio and 
far-infrared features being photo-ionized and photo-dissociated, 
respectively, by UV sources very close to Sgr~A$^*$.
\cite{latvakoski&sgh_99} also detected a far-infrared ring
running along the inner edge of the CNR, with radial thickness 
$\approx 8'' - 11''$ ($0.32~{\rm pc} - 0.44~{\rm pc}$), 
and intersecting the far-infrared minispiral at its western arc.\footnote{
To avoid any possible ambiguity, we reserve upper case (Minispiral, 
Northern and Eastern Arms, Western Arc) for the original radio features, 
and use lower case (minispiral, northern and eastern arms, western arc) 
for their far-infrared counterparts.
}
They naturally identified this far-infrared ring as 
the photo-dissociated inner layer of the CNR.
The masses of (presumably photo-dissociated) hydrogen traced 
in the far-infrared are 
$\approx 110~M_\odot$ in the northern arm,
$\approx 50~M_\odot$ in the eastern arm + bar,
$\approx 660~M_\odot$ in the western arc,
$\approx 1320~M_\odot$ in the full far-infrared ring
and $\approx 16~M_\odot$ in the Central Cavity outside the minispiral.
Note that the northern arm stretches $\approx 20''$ (0.8~pc) north 
beyond the CNR and that the $110~M_\odot$ mass refers only to its portion 
inside the CNR.
The mean hydrogen densities near the southwest and northeast ends 
of the far-infrared ring, as obtained from the hydrogen 
column densities divided by a line-of-sight depth of 1~pc, 
are $\approx 1.6 \times 10^4~{\rm cm}^{-3}$ 
and $\approx 4.0 \times 10^4~{\rm cm}^{-3}$, respectively.
Lastly, the dust temperature is found to be in the range $\approx (60-120)$~K.
\cite{latvakoski&sgh_99} proposed a simple, self-consistent dust model,
adjusted to reproduce the far-infrared data as well as possible.
In their model, the far-infrared ring (including the western arc)
is nearly circular, with inner radius 1.58~pc, 
axial thickness 0.4~pc and inclination $65^\circ$ to the plane of the sky.
The northern arm and the eastern arm + bar are both on parabolic orbits 
about Sgr~A$^*$, $10^\circ$ and $85^\circ$ out of the plane of the CNR,
respectively.
The (poorly constrained) mean hydrogen densities are
$\sim 1.2 \times 10^4~{\rm cm}^{-3}$ in the far-infrared ring,
$\sim 4 \times 10^4~{\rm cm}^{-3}$ in the core of the northern arm,
$\sim 2 \times 10^3~{\rm cm}^{-3}$ in the core of the eastern arm + bar
and $\sim 25~{\rm cm}^{-3}$ in the Central Cavity outside the minispiral.

The last gaseous component present inside the Central Cavity 
is a hot plasma, which was discovered with {\it Chandra} 
through its diffuse X-ray thermal emission \citep{baganoff&mmb_03}.
This hot X-ray emitting plasma extends across the central $\approx 20''$ 
(0.8~pc) of the Galaxy. 
A fit to its X-ray spectrum yields a temperature $\approx 1.3$~keV 
($1.5 \times 10^7$~K) and an electron density 
$\approx (26~{\rm cm}^{-3}) \, \phi_{\rm h}^{-1/2}$, 
where $\phi_{\rm h}$ is the hot plasma filling factor.
If the plasma is fully ionized and has twice the solar abundances,
its total mass is $\approx (0.1~M_\odot) \, \phi_{\rm h}^{1/2}$.
\cite{rockefeller&fmw04} showed with three-dimensional numerical
simulations that this hot plasma could be entirely explained as 
shocked gas produced in collisions between stellar winds.

\subsection{\label{obs_CNR}The Circumnuclear Ring}

The CNR has been observed in dust thermal continuum emission
\citep[e.g.][]{becklin&gw_82, mezger&zsw_89, davidson&wwl_92, dent&mwd_93, 
telesco&dw_96, latvakoski&sgh_99}
as well as in a wide variety of atomic and molecular tracers,
including the  21-cm line of H\,{\sc i} \citep{liszt&vbo_83},
fine-structure lines of [O\,{\sc i}] and [C\,{\sc ii}]
\citep{genzel&ctw_85, poglitsch&sgh_91},
and various transitions of 
H$_2$ \citep{gatley&jhw_84, gatley&jhw_86, depoy&gm_89, burton&a_92,
yusef&sbw_01},
CO \citep{liszt&bv_85, genzel&ctw_85, harris&jsg_85, serabyn&gww_86, 
gusten&gwj_87, sutton&djm_90, bradford&snb_05}, 
OH \citep{genzel&ctw_85},
CS \citep{serabyn&gww_86, serabyn&ge_89, montero&hh_09},
HCN \citep{gusten&gwj_87, marr&wb_93, jackson&ggr_93, marshall&lh_95, 
wright&cmh_01, christopher&ssy_05, montero&hh_09},
HCO$^+$ \citep{marr&wb_93, wright&cmh_01, shukla&ys_04, christopher&ssy_05},
NH$_3$ \citep{coil&h_99, mcgary&co_01, herrnstein&h_05}, etc.
Collectively, these tracers lead to the picture of an asymmetric, 
extremely clumpy, torus-like CNR, 
with a sharp inner boundary at radius $\approx (1 - 1.5)$~pc
and a more blurry, irregular outer boundary 
at radius $\approx (2.5 - 3)$~pc to the northeast 
and $\approx (4 - 7)$~pc to the southwest (see Figure~\ref{CNR}).
Besides, the CNR appears to be orbiting about Sgr~A$^*$ 
and to have considerable internal motions. 

Shortly after \cite{becklin&gw_82} discovered the CNR in dust
far-infrared continuum emission, \cite{genzel&ctw_85} observed it 
in several far-infrared emission lines, namely, in fine-structure lines
of [O\,{\sc i}] and [C\,{\sc ii}] and in rotational lines of CO and OH.
Their observations revealed a disk or torus of neutral gas
around the Central Cavity, with inner radius $\approx 1.4$~pc,
outer radius $\gtrsim 4.2$~pc (both rescaled to $r_\odot = 8.5$~kpc),
inclination $\approx 69^\circ$ to the plane of the sky,
and tilt $\approx 20^\circ$ to the Galactic plane.
Atomic and molecular gases were found to be mixed throughout the disk/torus, 
with the fraction of atomic gas decreasing outward
-- as expected for a photo-dissociation region illuminated from inside. 
From the intensities of the [O\,{\sc i}] and [C\,{\sc ii}] lines,
\cite{genzel&ctw_85} estimated the temperature of the atomic gas 
at $\approx 300$~K and its true hydrogen density at 
$\approx 10^5~{\rm cm}^{-3}$.
Furthermore, from existing dust far-infrared continuum emission data, 
they derived a space-averaged hydrogen density 
$\approx 7 \times 10^3~{\rm cm}^{-3}$ in the [O\,{\sc i}] emission region
(assuming a size $\approx (2-3)$~pc) 
and a total hydrogen mass $\approx 1.5 \times 10^4~M_\odot$ 
within a radius of 4.2~pc,
while from existing CO $J = 1 \to 0$ line emission data, 
they derived a total hydrogen mass $\approx (1.5-3.7) \times 10^4~M_\odot$ 
in the purely molecular gas beyond 4.2~pc
(all masses were rescaled to $r_\odot = 8.5$~kpc).

Subsequent CO and CS observations offered additional insight
into the physical conditions of the CNR.
\cite{harris&jsg_85} used their 0.37~mm CO $J = 7 \to 6$ line emission 
measurements in conjunction with previous measurements of two lower 
and two higher CO rotational lines to determine the H$_2$ density 
and gas temperature in CO-emitting regions.
They obtained best-fit values $\approx 3 \times 10^4~{\rm cm}^{-3}$ 
and $\approx 300$~K, respectively, and they restricted the range of
acceptable density-temperature combinations to 
$(5 \times 10^5~{\rm cm}^{-3}, 150~{\rm K})$
$-$ $(10^4~{\rm cm}^{-3}, 450~{\rm K})$.
They also concluded that the CO-emitting gas is very clumpy, 
with a volume filling factor $\sim 0.1$.
\cite{serabyn&gww_86}, who observed the CNR in 2.6~mm CO $J = 1 \to 0$ 
and 3.1~mm CS $J = 2 \to 1$ emission, inferred densities 
of a few $10^5~{\rm cm}^{-3}$ for the CS-emitting gas, 
and derived a mass of molecular gas in the CNR 
$\gtrsim 1.5 \times 10^4~M_\odot$ (rescaled to $r_\odot = 8.5$~kpc) 
from the measured CO flux.
\cite{sutton&djm_90} combined their 0.87~mm CO $J = 3 \to 2$ observations 
with existing CO $J = 1 \to 0$ and $J = 7 \to 6$ data 
to find that the H$_2$ density and gas temperature in CO-emitting regions 
vary from $\approx 2 \times 10^5~{\rm cm}^{-3}$ 
and $\approx 200$~K in the inner parts of the CNR 
to $\approx 2 \times 10^4~{\rm cm}^{-3}$ and $\approx 100$~K in the outer parts.
They also confirmed the clumpiness of the CO-emitting gas 
and estimated its volume filling factor at $\sim 0.05$.
The more recent CO $J = 7 \to 6$ observations of \cite{bradford&snb_05},
which the authors analyzed together with published data on two lower 
and two higher CO rotational transitions, 
taking radiative transfer into account,
yielded an H$_2$ density $\approx (5-7) \times 10^4~{\rm cm}^{-3}$ 
and a gas temperature $\approx (200-300)$~K.

Two other frequently used diagnostic molecules are HCN and HCO$^+$.
In the 3.4~mm HCN $J = 1 \to 0$ emission map of \cite{gusten&gwj_87},
the CNR emerges as an almost complete molecular ring centered $\approx 8''$ 
(0.32~pc) southeast of Sgr~A$^*$, 
which has projected major and minor mean diameters
$\approx 95''$ and $50''$ (4.0~pc and 2.1~pc), respectively.
The major axis has a position angle $\approx 30^\circ$ east of north,
so that it is approximately aligned with the Galactic plane
(see Figure~\ref{axes}).
Moreover, if the ring is intrinsically circular, the aspect ratio 
$\approx 2\!:\!1$ implies an inclination angle $\approx 60^\circ$ 
out of the plane of the sky.
The ring's inner and outer radii along the major axis are 
$\approx 30''$ and $65''$ (1.2~pc and 2.7~pc), respectively,
but on the southwest side the HCN emission extends out to 
$\approx 100''$ (4.2~pc)
-- for comparison, \cite{serabyn&gww_86} derived an outer radius
$\approx 7$~pc for CO $J = 1 \to 0$ emission on the southwest side.
The axial thickness of the whole HCN structure increases steadily 
with radius from $\approx 0.42$~pc at 1.7~pc to $\approx 1.2$~pc at 4.2~pc
(rescaled to $r_\odot = 8.5$~kpc).
\citeauthor{gusten&gwj_87}'s (\citeyear{gusten&gwj_87}) study 
also provides important kinematic information.
The measured radial velocity peaks at $\simeq 100~{\rm km~s}^{-1}$,
and its variation with position angle on the sky agrees reasonably well
with that expected for rotation of a warped ring with rotation velocity 
$\approx 110~{\rm km~s}^{-1}$ \!/\! $135~{\rm km~s}^{-1}$
and inclination angle $\approx 70^\circ$ \!/\! $50^\circ$
in the southwest \!/\! northeast parts.
In addition to this overall rotation, the gas exhibits strong turbulent 
motions with a velocity dispersion decreasing from an average of 
$\approx 55~{\rm km~s}^{-1}$ near the inner edge
to $\approx 37~{\rm km~s}^{-1}$ near the southwest outer edge.

\cite{jackson&ggr_93}, who mapped a somewhat smaller area 
in the 1.1~mm HCN $J = 3 \to 2$ emission line, reached slightly 
different conclusions.
They obtained a velocity-integrated HCN map similar to that of 
\cite{gusten&gwj_87}, but they interpreted the kinematic data 
in a different manner.
Instead of invoking a single rotating ring that is warped,
they appealed to four separate kinematic components,
the most prominent of which is a rotating circular ring 
of peak radius $\approx (1.5 - 2)$~pc,
inclination angle $\approx 65^\circ - 75^\circ$,
position angle of the projected major axis $\approx 25^\circ$ east of north
and rotation velocity $\approx 110~{\rm km~s}^{-1}$.
The other, weaker components are the so-called
$50~{\rm km~s}^{-1}$ Streamer, $70~{\rm km~s}^{-1}$ Feature
and $-20~{\rm km~s}^{-1}$ Cloud.
For the physical parameters of the molecular gas in the CNR,
model calculations of HCN excitation and radiative transport yielded 
a temperature $\approx (50 - 200)$~K,
a true H$_2$ density $\sim (10^6 - 10^8)~{\rm cm}^{-3}$
and a space-averaged H$_2$ density $\sim (10^4 - 10^5)~{\rm cm}^{-3}$
(assuming a line-of-sight pathlength of 1~pc through the CNR).

\cite{wright&cmh_01} imaged the central 12~pc simultaneously 
in the 3.4~mm HCN and HCO$^+$ $J = 1 \to 0$ transitions, 
both of which are supposed to trace molecular gas 
with density $\sim (10^5 - 10^6)~{\rm cm}^{-3}$.
The two tracers present essentially the same velocity-integrated emission,
and both indicate that the CNR is not a disk, but a well-defined ring
peaked at radius $\approx 45''$ (1.9~pc) that extends radially 
from $\approx 20''$ to $60''$ (0.8~pc to 2.5~pc), 
with a southwest extension out to $\approx 120''$ (5~pc).
Kinematically, the CNR is found to consist of two or three distinct 
streamers rotating around Sgr~A$^*$ on separate orbits.
The different inclinations of these orbits would be the reason 
for the warped-ring appearance of the CNR.

\cite{christopher&ssy_05} performed additional HCN and HCO$^+$ 
$J = 1 \to 0$ observations with enhanced spatial resolution, 
which enabled them to study the internal structure of the CNR 
in greater detail.
Their velocity-integrated maps are on the whole similar to those of
\cite{wright&cmh_01}, and they, too, display a well-defined ring, 
although with slightly different dimensions.
Here, the azimuthally-averaged HCN emission is found to peak at radii 
between $\approx 40''$ and $50''$ (1.7~pc and 2.1~pc), 
drop off steeply (over $\approx 10''$) on either side of the peak, 
and then decline much more gradually out past $\approx 150''$ (6.2~pc).
\cite{christopher&ssy_05} were able to resolve 26 dense molecular cores 
within the CNR, with typical sizes $\approx 7''$ (0.3~pc), and estimated
their masses in two independent manners:
virial masses were derived from the measured sizes and velocity widths, 
assuming the cores to be gravitationally bound, 
and optically-thick masses were derived from the measured sizes 
and HCN column densities, assuming an HCN-to-H$_2$ ratio of $10^{-9}$
\citep[as opposed to $2 \times 10^{-8}$ in][]{jackson&ggr_93}
and multiplying by 1.36 to account for helium.
Both masses were found to agree well, with median values 
$\simeq 1.7 \times 10^4~M_\odot$ and $\simeq 2.4 \times 10^4~M_\odot$,
respectively, corresponding to mean H$_2$ densities inside the dense cores 
$\approx 4 \times 10^7~{\rm cm}^{-3}$ and $\approx 5 \times 10^7~{\rm cm}^{-3}$.
From their derived core masses, \cite{christopher&ssy_05} estimated 
the total mass of the CNR at $\approx 10^6~M_\odot$.

Remarking that the HCN and HCO$^+$ $J = 1 \to 0$ emission lines 
from the GC region are subject to self-absorption
due to the intervening (cooler and more diffuse) molecular gas,
\cite{montero&hh_09} turned to the higher 0.85~mm HCN $J = 4 \to 3$ 
transition, which they observed toward the CNR,
along with the 0.87~mm CS $J = 7 \to 6$ transition. 
They detected very clumpy emission from both tracers,
with clumps arranged in a necklace-like fashion around the CNR.
The southern part of the CNR has stronger emission
and is found to be denser and warmer (from a comparison with 
the previously measured HCN $J = 1 \to 0$ line) than the northern part.
Similarly, the inner edge of the CNR appears to be warmer than 
the outer parts -- as expected if the CNR is heated by the intense
UV radiation from the central star cluster.
The clumps present wide ranges of physical characteristics,
with sizes $\approx 3'' -13''$ ($0.12~{\rm pc} - 0.5~{\rm pc}$),
virial masses $\simeq (0.4-60) \times 10^4~M_\odot$
and virial H$_2$ densities $\approx (2-40) \times 10^7~{\rm cm}^{-3}$.
Summing the virial masses of all the HCN $(4 \! \to \! 3)$ clumps 
listed by \cite{montero&hh_09} gives a total CNR mass 
$\approx 1.3 \times 10^6~M_\odot$, comparable to the CNR  mass 
estimated by \cite{christopher&ssy_05}.

The physical parameters of the CNR were very recently re-estimated 
by \cite{oka&nkt_11}, based on several millimeter and submillimeter 
molecular emission lines, including the $J = 1 \to 0$ lines of CO, HCN, 
HCO$^+$, N$_2$H$^+$, HNC and SiO, the $J = 2 \to 1$ line of SiO 
and the $J = 3 \to 2$ line of CO.
A one-zone large-velocity-gradient radiative-transfer analysis 
of a restricted set of lines
(CO $J = 1 \to 0$, CO $J = 3 \to 2$ and HCN $J = 1 \to 0$),
assuming [CO]/[H$_2$] = $2.4 \times 10^{-5}$
and [HCN]/[H$_2$] = $4.8 \times 10^{-8}$,
concludes that the bulk of the CNR is made of molecular gas 
with kinetic temperature $\gtrsim 40$~K and H$_2$ density 
$\approx (5 \times 10^3 - 2 \times 10^4)~{\rm cm}^{-3}$, 
the best-fit values being 63~K and $1.26 \times 10^4~{\rm cm}^{-3}$, 
respectively.
Comparisons between the various line-intensity maps (most importantly,
the CO and HCN $J = 1 \to 0$ maps) show that the innermost ring, 
at radius $\approx 2$~pc, is both warmer and denser than the bulk of the CNR.
The total H$_2$ mass of the CNR, estimated from the $^{13}$CO $J = 1 \to 0$ 
intensity map, is $\approx (2.3-5.2) \times  10^5~M_\odot$,
which corresponds to the typical mass of GMCs in the GC region.
Much larger is the virial mass of the CNR, estimated at
$\approx 5.7 \times 10^6~M_\odot$.
According to the authors, the important discrepancy between both masses 
strongly suggests that the CNR is not bound by self-gravity,
but rather by the central mass.
Finally, the CO $J = 3 \to 2$ data, interpreted with a simple 
kinematic model, point to a two-regime situation, where the bulk of the CNR 
is infalling at a speed $\approx 50~{\rm km~s}^{-1}$,
while the innermost ring at $\approx 2$~pc is predominantly rotating.

Numerical simulations have greatly contributed to enhance 
our understanding of the CNR.
For instance, the sticky-particle calculations of \cite{sanders_98} 
showed that the morphology and kinematics of the CNR 
could be explained by the tidal capture and disruption of 
a low angular-momentum cloud by the central massive black hole.
The cloud would first be stretched into a long filament,
which would wrap about the dynamical center and collide a few times 
with itself.
Under the effect of viscous dissipation, the tidal debris would then 
settle into an asymmetric, precessing dispersion ring, 
which would persist for $\gtrsim 10^6$~yr.
A similar scenario could apply to the Northern Arm
(with its westward extension) inside the Central Cavity,
although the original cloud would have to be smaller and 
on a lower angular-momentum orbit.
It is interesting that, in the best-fitting simulation, 
the orbital plane of the extended Northern Arm lies at $10^\circ$
of that of the CNR -- which agrees well with the findings of
\cite{latvakoski&sgh_99}, \cite{paumard&mm_04} and \cite{zhao&bmd_10}
(see Section~\ref{obs_cavity}).

\subsection{\label{obs_shell}The Sgr~A East SNR}

The nonthermal radio source Sgr~A East clearly has a shell-like structure.
The VLA 20~cm radio continuum map of \cite{ekers&gsg_83} shows that 
this shell is elongated along the Galactic plane,
with projected dimensions $\approx 3\fmn6 \times 2\fmn7$
($9.0~{\rm pc} \times 6.7~{\rm pc}$),
and that it is off-centered by $\approx 2.1$~pc slightly north of east 
from Sgr~A$^*$.
In projection, the western side of the Sgr~A East shell 
overlaps with the Sgr~A West H\,{\sc ii} region,
and the Western Arc appears to smoothly merge into the shell.
The shell-like morphology of Sgr~A East, its measured size
and its nonthermal (supposedly synchrotron) radio emission 
all converge to indicate that it is an SNR
-- as initially proposed by \cite{jones_74} and \cite{ekers&gsd_75}.

In the VLA 90~cm radio continuum image of \cite{pedlar&aeg_89},
the Sgr~A East shell has projected dimensions $\approx 3\fmn3 \times 2\fmn1$ 
($8.2~{\rm pc} \times 5.2~{\rm pc}$)
and its major axis is at position angle $\approx 40^\circ$ east of north,
i.e., roughly parallel to the Galactic plane (see Figure~\ref{axes}).
On the western side of the radio shell, the spiral pattern of the Sgr~A West 
H\,{\sc ii} region clearly stands out in absorption (free-free absorption 
by thermal gas) against the nonthermal emission from Sgr~A East.
This definitely places Sgr~A West in front, or close to the near surface,
of Sgr~A East -- as argued before by \cite{yusef&m_87a}, based on 
a comparison of 6~cm and 20~cm radio continuum maps.
Yet not all of the 90~cm emission from Sgr~A East is actually absorbed
in this direction, which may indicate that Sgr~A West lies within
Sgr~A East and close to its near surface 
(\citeauthor{yusef&mw_00} \citeyear{yusef&mw_00}; 
see also \citeauthor{maeda&bfm_02} \citeyear{maeda&bfm_02}).
On the eastern side, the boundary of the radio shell is strikingly straight, 
which suggests that Sgr~A East is colliding with 
the neighboring M$-$0.02$-$0.07 GMC.

Additional support for this suggestion comes from the 1.3~mm observations
of \cite{mezger&zsw_89}, which reveal a dust ring surrounding 
the Sgr~A East radio shell.
This dust ring, with major inner diameter $\approx 10$~pc along 
the Galactic plane, was also detected in OH and H$_2$CO absorption 
\citep{sandqvist_74, whiteoak&rl_74}
as well as in CO emission \citep[see ][]{mezger&dz_96}.
Its eastern part coincides with a ridge of dense molecular gas
in M$-$0.02$-$0.07 (see Section~\ref{obs_clouds}) and its southern part 
coincides with dense molecular gas belonging to M$-$0.13$-$0.08.
\cite{mezger&zsw_89} interpreted the observed ring as a shell 
of gas and dust surrounding the radio shell that is produced by 
the same supernova explosion.
The reason why the gas-and-dust shell is particularly dense toward
the east and south would be because in these directions the SNR 
has expanded into the M$-$0.02$-$0.07 and M$-$0.13$-$0.08 GMCs,
respectively.
Toward the west, the shell may have encountered Sgr~A$^*$ and 
be captured (at least partially) by its gravitational pull, 
so as to form the present-day CNR.
The authors also pointed out that the explosion seems to have dispersed
most of the gas in front of Sgr~A West,
which could be explained if the explosion occurred inside a GMC, 
close to its near surface. 
The total hydrogen mass swept-up into the gas-and-dust shell 
was estimated at $\approx 6 \times 10^4~M_\odot$, which implies 
a mean preshock density $\sim 10^4~{\rm cm}^{-3}$ in the parent GMC.

The detection of 1720~MHz OH masers, 
without 1665~MHz and 1667~MHz counterparts, 
along the periphery of Sgr~A East \citep{yusef&rgf_96, yusef&rgf_99}
revealed the presence of shocked molecular gas,
thereby providing independent evidence 
that the expanding SNR is interacting with nearby molecular clouds.
At the southeastern boundary of Sgr~A East with the M$-$0.02$-$0.07 cloud,
1720~MHz OH masers were detected with radial velocities between
$49~{\rm km~s}^{-1}$ and $66~{\rm km~s}^{-1}$,
i.e., close to the $50~{\rm km~s}^{-1}$ systemic velocity of the cloud,
which reinforces the concept that the Sgr~A East shock is propagating 
into M$-$0.02$-$0.07.
1720~MHz OH masers were also detected toward the CNR,
with radial velocities of $134~{\rm km~s}^{-1}$ 
(at the intersection between the Northern Streamer and the CNR)
and $43~{\rm km~s}^{-1}$ (along the outer western edge of the CNR).
To confirm the presence of shocked molecular gas near the OH masers,
\cite{yusef&sbw_01} looked for $2.12~\mu$m H$_2$ $v=1-0 \ S(1)$ emission, 
and they found that all but one of the OH masers detected in the region 
are indeed accompanied by H$_2$ emission. 
In particular, the $43~{\rm km~s}^{-1}$ OH maser lies in projection 
along an H$_2$ filament, which extends over $\approx 1'$ 
along the western boundary of Sgr~A East
and peaks at velocities $\approx (50-75)~{\rm km~s}^{-1}$,
close to the peak velocities of the western edge of the CNR.
The location, morphology and kinematics of the H$_2$ filament
and its likely association with a 1720~MHz OH maser
strongly suggest that the filament was generated by the passage of 
the Sgr~A East shock over the CNR \citep{yusef&sbw_99}.
This, in turn, implies that Sgr~A East must have engulfed part of the CNR. 

Inside the cavity of the Sgr~A East radio shell, \cite{maeda&bfm_02}
observed a hot X-ray emitting plasma with {\it Chandra}.
They noted that the X-ray emission is concentrated within the central 
$\approx 2$~pc of the radio shell.
From the measured X-ray spectrum (continuum + K$\alpha$ emission lines
from highly ionized metals), they inferred
a temperature $\approx 2.1$~keV ($2.4 \times 10^7$~K) and
an overabundance of heavy elements by a factor $\approx 4$ with respect
to solar levels, with an inward gradient in the abundance of iron
relative to the other metals.
Assuming a spherical volume of radius 1.6~pc, they derived 
an electron density $\approx (6~{\rm cm}^{-3}) \, \phi_{\rm h}^{-1/2}$,
a total gas mass $\approx (2~M_\odot) \, \phi_{\rm h}^{1/2}$
and a total thermal energy 
$\approx (2 \times 10^{49}~{\rm ergs}) \, \phi_{\rm h}^{1/2}$,
where again $\phi_{\rm h}$ is the hot plasma filling factor.
This estimated gas mass and thermal energy, together with the strong
enrichment in heavy elements, lends credence to the long-standing idea 
that Sgr~A East is an SNR.
Moreover, the combination of shell-like nonthermal radio emission
and centrally concentrated X-ray thermal emission classify this SNR
as a mixed-morphology SNR.

\cite{sakano&wdp_04} obtained a higher-quality X-ray spectrum of Sgr~A East
with {\it XMM-Newton}.
Both their spectral fitting and their line-ratio analysis require 
at least two temperature components, at $\approx 1$~keV 
and $\approx 4$~keV, respectively.
The derived temperatures are somewhat lower in the core of the X-ray source
($\approx 0.9$~keV and $\approx 3$~keV, respectively).
The Fe abundance varies from $\approx 4$ times solar in the core 
down to $\approx 0.5$ solar in the outer region,
whereas other metals (S, Ar, Ca) have more uniform abundances, 
all in the range $\approx (1-3)$ solar.
If the core is approximated as a $28''$ (1.1~pc) radius sphere,
and if the low- and high-temperature components within it 
are in thermal pressure balance and have a combined filling factor 
$\phi_{\rm h}$, their respective electron densities are 
$\approx (20~{\rm cm}^{-3}) \, \phi_{\rm h}^{-1/2}$
and $\approx (6~{\rm cm}^{-3}) \, \phi_{\rm h}^{-1/2}$.
The corresponding total mass and thermal energy of hot plasma in the core 
are $\approx (1.4~M_\odot) \, \phi_{\rm h}^{1/2}$
and $\approx (1.3 \times 10^{49}~{\rm ergs}) \, \phi_{\rm h}^{1/2}$,
with $65\%$ of the mass and $38\%$ of the thermal energy 
in the low-temperature component.
The distinct overabundance of Fe in the core (and not outside) suggests 
that the above estimates refer to stellar ejecta, which is consistent 
with a single supernova explosion.
The rest of the X-ray emitting plasma is more likely shocked 
interstellar matter.

Much deeper {\it Chandra} observations than those of \cite{maeda&bfm_02} 
enabled \cite{park&mbm_05} to perform a spatially resolved spectral 
analysis of Sgr A East.
They observed enhanced hard X-ray emission from a Fe-rich plasma
over a $\approx 40''$ (1.7~pc) diameter region near the center of the SNR.
They naturally identified this bright, Fe-rich plasma with stellar ejecta.
Like \cite{sakano&wdp_04}, they fitted its hard X-ray spectrum with
two temperatures (estimated here at $\approx 1$~keV and $\approx 5$~keV)
and they derived a high Fe abundance ($\approx 6$ times solar) 
compared to the S, Ar, Ca abundances ($\approx (0.7-1.8)$ solar).
\cite{park&mbm_05} also observed soft X-ray emission outside 
the hard X-ray core, in particular, in a plume-like feature 
extending toward the north of the SNR.
They found that the emitting plasma in this feature could be characterized 
by a single temperature ($\approx 1.3$~keV) and solar abundances, 
and they identified it with shocked interstellar matter.
\cite{park&mbm_05} also provided density estimates,
both in the central Fe-rich core and in the northern plume-like feature.
For the latter, they adopted a half-conical volume with a circular base 
of radius $\approx 25''$ and a height $\approx 50''$, and they obtained
an electron density $\approx (7.4~{\rm cm}^{-3}) \, \phi_{\rm h}^{-1/2}$.
For the central core, they assumed a $\approx 40''$ diameter sphere
with pure Fe$^{24+}$ ejecta, and they obtained electron densities
$\approx (2.3~{\rm cm}^{-3}) \, \phi_{\rm h}^{-1/2}$
and $\approx (0.5~{\rm cm}^{-3}) \, \phi_{\rm h}^{-1/2}$
in the low- and high-temperature components, respectively, 
while they derived a total Fe ejecta mass 
$\approx (0.15~M_\odot) \, \phi_{\rm h}^{1/2}$.

Finally, with {\it Suzaku}, \cite{koyama&uhm_07} acquired a detailed 
X-ray spectrum of Sgr~A East, which displays all the previously 
(firmly or tentatively) reported emission lines
(K$\alpha$ lines from He-like S, Ar, Ca, Fe;
K$\alpha$ lines from H-like S, Ar, Fe)
as well as a number of newly discovered emission lines
(K$\alpha$ line from He-like Ni;
K$\beta$ lines from He-like S, Ar, Fe;
K$\gamma$ line from He-like Fe).
The measured line ratios confirmed the necessary presence 
of at least two temperature components,
while the complete spectral fitting required an additional hard tail.
Altogether, the best-fit spectrum consists of two thin thermal 
components, with $\approx 1.2$~keV and $\approx 6$~keV,
plus a power-law component, which could be caused by either a collection 
of point sources or non-thermal clumps and filaments.
\cite{koyama&uhm_07} found that, on average over the SNR, S, Ar, Ca
have roughly solar abundances, while Fe is overabundant by a factor 
$\approx 2-3$,
and they estimated the total mass of hot plasma
at $\approx (27~M_\odot) \, \phi_{\rm h}^{1/2}$.

\subsection{\label{obs_halo}The radio halo}

The Sgr~A East shell appears to be surrounded by an extended radio halo.
In the VLA 20~cm radiograph of \cite{yusef&m_87a}, the radio halo
has approximately the same shape (roughly elliptical),
aspect ratio ($\sim 1.5$), orientation (parallel to the Galactic plane)
and center (northeast of Sgr~A$^*$) as Sgr~A East,
but it is about twice as large ($\sim 20$~pc along its major axis).
These properties suggest that Sgr~A East and the radio halo are part 
of the same physical system.
One possibility would be that the radio halo results from a leakage 
of cosmic-ray electrons accelerated in the Sgr~A East SNR.

\cite{pedlar&aeg_89} obtained more information on the nature and
physical characteristics of the radio halo by combining VLA 90~cm, 
20~cm and 6~cm continuum observations of the Sgr~A complex.
The radio halo is clearly visible in the 90~cm image, where it has 
a roughly triangular shape, with a total extent $\approx 7'$ (17.5~pc).
The entire Sgr~A East shell shows a low-frequency turnover
in its nonthermal emission,
which can be explained by free-free absorption 
by thermal ionized gas with an emission measure 
$\approx (1-2) \times 10^5~{\rm pc~cm}^{-6}$ 
(assuming an electron temperature $\approx 5\,000$~K).
\cite{pedlar&aeg_89} suggested that the absorbing thermal gas 
belongs to the $7'$ radio halo.
The radio halo itself has mainly nonthermal emission
(at the considered wavelengths), and it, too, shows a low-frequency 
turnover explainable by free-free absorption.
However, here, instead of residing in a separate foreground screen, 
the absorbing thermal gas is more probably mixed with the emitting
nonthermal gas within the halo.
In other words, the $7'$ radio halo is likely to comprise a mixture 
of thermal and nonthermal gases.

\cite{pedlar&aeg_89} were able to reproduce the spectrum 
of the radio halo by adopting for the thermal gas
an electron temperature $\simeq 5\,000$~K,
an emission measure $\simeq 2.7 \times 10^5~{\rm pc~cm}^{-6}$,
and {\it FWHM} dimensions $\simeq 4' \times 4'$ 
($10~{\rm pc} \times 10~{\rm pc}$),
which, for a spherical distribution and a filling factor of unity, 
imply an electron density $\simeq 165~{\rm cm}^{-3}$
and an H$^+$ mass $\simeq 2\,100~M_\odot$
(rescaled to $r_\odot = 8.5$~kpc).
Furthermore, since the thermal-gas free-free optical depths required 
to explain the low-frequency turnovers of the Sgr~A East shell and
of the radio halo are similar, \cite{pedlar&aeg_89} suggested that 
the radio halo is mostly situated in front of Sgr~A East.
It should be noted, however, that within the uncertainties, 
the derived free-free optical depths are also consistent with 
only the near half of the radio halo lying in front of Sgr~A East,
so that Sgr~A East could actually be deeply embedded within the radio halo,
and even concentric with it.

The presence of warm ionized gas in the radio halo is confirmed 
by observations of radio recombination lines, which, in addition, 
provide useful kinematic information.
VLA observations of the 1375~MHz (22~cm) H168$\alpha$ recombination line
by \cite{anantharamaiah&pg_99} revealed an extended area of 
H168$\alpha$ emission encompassing the entire Sgr~A East shell 
and covering a broad range of radial velocities
from $\approx -200~{\rm km~s}^{-1}$ to $\approx +50~{\rm km~s}^{-1}$.
The fact that the ionized gas observed in the H168$\alpha$ line
is detected neither in the lower-frequency H270$\alpha$ line
(sensitive to $n_{\rm e} \lesssim 10~{\rm cm}^{-3}$)
nor in the higher-frequency H110$\alpha$ and H92$\alpha$ lines
(sensitive to $n_{\rm e} \gtrsim 1\,000~{\rm cm}^{-3}$)
constrains the electron density to lie in the range
$n_{\rm e} \sim (10 - 1\,000)~{\rm cm}^{-3}$.
The electron density can be additionally constrained
by considering the H168$\alpha$ data in conjunction with the radio
spectrum of Sgr~A East obtained by \cite{pedlar&aeg_89}
and by assuming that the H168$\alpha$ emission arises in the same 
thermal ionized gas as the free-free absorption that is responsible for 
the low-frequency turnover of Sgr~A East.
In this manner, \cite{anantharamaiah&pg_99} found that a model with
electron temperature $\approx 10^4$~K,
emission measure $\approx 3.3 \times 10^5~{\rm pc~cm}^{-6}$
and electron density $\approx 100~{\rm cm}^{-3}$
gave a good fit to all the data combined.
The H$^+$ mass predicted by this model is $\sim 8 \times 10^4~M_\odot$
(this is the value quoted by the authors, but we believe they meant 
an H$^+$ mass $\sim 8 \times 10^3~M_\odot$)
over the $\sim 4' \times 4'$ projected area of Sgr~A East.

\cite{maeda&bfm_02} suggested that the halo of ionized gas 
roughly corresponds to the region of non--solid-body rotation 
around Sgr~A$^*$.
This region would have a relatively homogeneous density,
because differential rotation would have sheared and smoothed out
the interstellar gas on a short ($\sim 10^5$~yr) timescale.
Regarding the source of ionization, \cite{maeda&bfm_02} ruled out
collisional ionization, which would require too high a temperature.
Instead, they argued in favor of photo-ionization by X rays,
and they proposed that the ionizing X rays were emitted by Sgr~A$^*$
$\sim (10^2 - 10^3)$~yr ago, during an episod of intense nuclear activity.
This episod could have been triggered by the passage over Sgr~A$^*$ 
of the gas-and-dust shell compressed by the Sgr~A East forward shock.
If this scenario is correct, Sgr~A$^*$ should presently reside inside 
the Sgr~A East cavity, consistent with Sgr~A West itself residing inside 
Sgr~A East \citep[see][]{yusef&mw_00}.

Although the existence of a radio halo around Sgr~A East leaves 
virtually no doubt, the presence of warm ionized gas within it 
is not universally accepted.
For instance, the idea was called into question by \cite{roy&r_09},
who measured the total flux densities of Sgr~A East and the radio halo 
at five different frequencies ranging from 154~MHz (195~cm) 
to 1.4~GHz (21~cm).
They observed similar low-frequency turnovers (at $\sim 400$~MHz) 
in the radio spectra of both sources, 
which they argued could be entirely attributed to free-free absorption
in a common foreground screen, without requiring the presence
of warm ionized gas inside the radio halo.
From this, they concluded that the radio halo is in fact
a purely nonthermal source.

\subsection{\label{obs_clouds}The belt of molecular clouds}

\begin{figure}[!t]
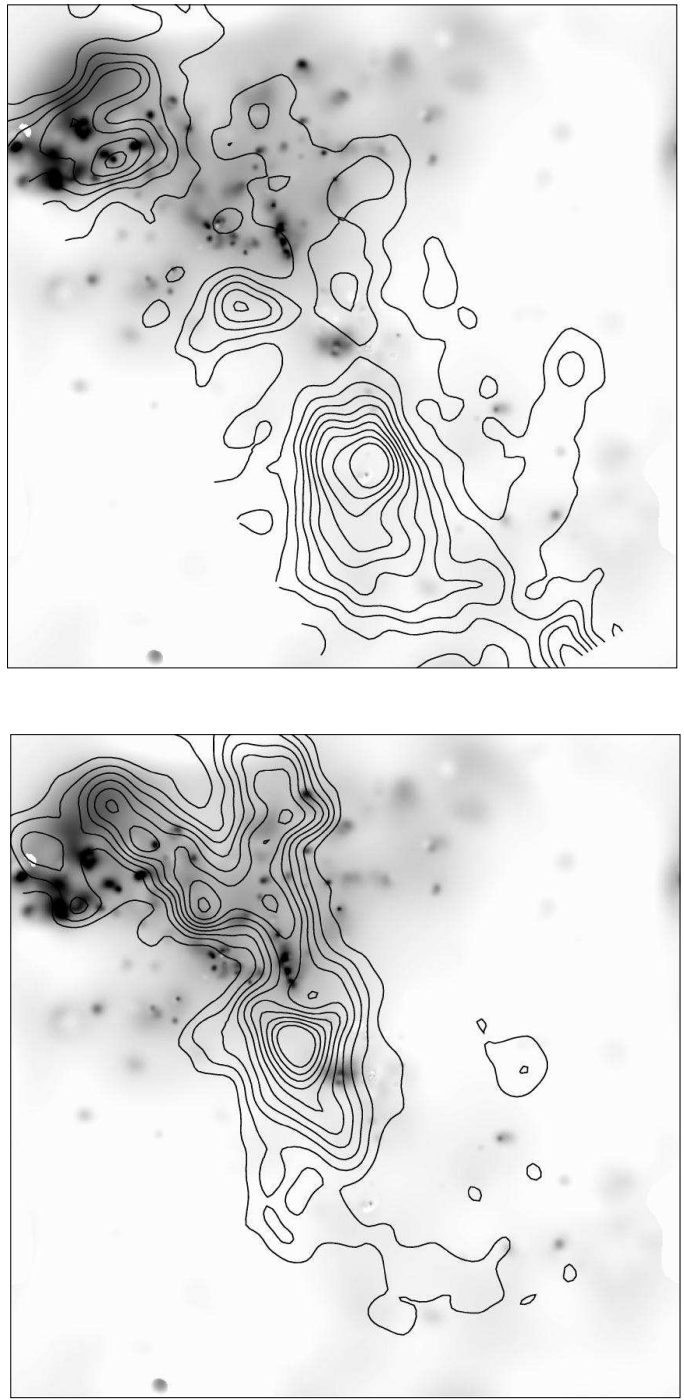

\centering
\includegraphics[width=0.48\textwidth]{sangwook_4c.epsi}
\bigskip\bigskip\\
\centering
\includegraphics[width=0.48\textwidth]{sangwook_4e.epsi}
\caption{
\label{CS}
Contour lines of the 6.1~mm CS $J = 1 \to 0$ line emission 
in the velocity ranges $(10-30)~{\rm km~s}^{-1}$ (top)
and $(40-50)~{\rm km~s}^{-1}$ (bottom),
superimposed on a grayscale equivalent-width image 
of the 6.4~keV low-ionization Fe K$\alpha$ line emission,
in a $17' \times 17'$ field of view centered on Sgr~A$^*$.
The CS data are from \cite{tsuboi&hu_99}
and the 6.4~keV data from \cite{park&mbm_04}.
Figure credit: Sangwook Park.
}
\end{figure}

The geometry, kinematics and physical state of molecular clouds 
around the Sgr~A complex have been investigated mainly 
through radio spectral lines of different molecules, including
CO \citep{solomon&spw_72},
NH$_3$ \citep{guesten&wp_81, okumura&ifc_89, okumura&ifh_91, 
coil&h_99, coil&h_00, mcgary&co_01, herrnstein&h_02, herrnstein&h_05},
CS \citep[][see Figure~\ref{CS}]{serabyn&la_92, tsuboi&hu_99, tsuboi&om_06, 
tsuboi&mo_09},
H$_2$ \citep{lee&pdh_03, lee&pcd_08},
CH$_3$OH \citep{stankovic&sml_07},
HC$_3$N \citep{sandqvist&lhb_08},
SiO \citep{amo&mmm_09, amo&mm_11}, etc.,
and also through dust submm continuum emission
\citep[e.g.,][]{mezger&zsw_89, zylka&mw_90, dent&mwd_93, lis&c_94}.
The early CO emission map of \cite{solomon&spw_72} already revealed
two massive molecular clouds peaking $\approx 3'$ east 
and $\approx 2\fmn5$ south of Sgr~A$^*$ and having radial velocities 
in the range $\approx (45 - 65)~{\rm km~s}^{-1}$ 
and $\approx (15 - 35)~{\rm km~s}^{-1}$, respectively.
\cite{solomon&spw_72} estimated their diameters at $\sim 6' - 20'$ 
and their hydrogen masses at $\gtrsim 10^5~M_\odot$.
Later, \cite{guesten&wp_81} carried out NH$_3$ observations of the region
and derived a fairly uniform gas temperature $\approx (50 - 120)$~K 
throughout the clouds.
They also labeled the clouds M$-$0.02$-$0.07 and M$-$0.13$-$0.08, 
respectively, according to the Galactic coordinates of their NH$_3$ 
emission peaks.
Today, these clouds are often referred to as the $50~{\rm km~s}^{-1}$ and 
$20~{\rm km~s}^{-1}$ clouds, respectively, although both denominations 
are not necessarily strictly equivalent.
For instance, some authors include into the $50~{\rm km~s}^{-1}$ cloud
not only M$-$0.02$-$0.07, but also several molecular knots
on its positive-longitude side.

\cite{zylka&mw_90} made a first important attempt to obtain a coherent
three-dimensional picture of molecular clouds in the central $\sim 50$~pc.
To locate gas along the line of sight, they compared their 1.3~mm dust
emission map to a $2.2~\mu$m intensity map showing dust absorption 
against the emission from the central star cluster,
and to determine the gas kinematics, they resorted to isotopic CO and CS 
spectroscopy.
In this manner, they found that M$-$0.02$-$0.07 is actually composed 
of two separate clouds, which they designated the Sgr~A East Core 
and the Curved Streamer. 
These two clouds, together with M$-$0.13$-$0.08, 
were found to have the following properties:

\noindent
$\bullet$ M$-$0.13$-$0.08, situated south of Sgr~A East, 
has projected dimensions $\approx 15~{\rm pc} \times 7.5~{\rm pc}$
(with major axis roughly parallel to the Galactic plane),
hydrogen mass $\approx 3 \times 10^5~M_\odot$ 
and radial velocities increasing with Galactic longitude
from $\approx 5~{\rm km~s}^{-1}$ to $\approx 25~{\rm km~s}^{-1}$.
Along the line of sight, the cloud lies in front of Sgr~A$^*$
\citep[as already pointed out by][based on H$_2$CO absorption]{guesten&d_80},
at a possible distance $\sim 50$~pc.

\noindent
$\bullet$ The Sgr~A East Core, which surrounds the Sgr~A East radio shell,
is $\gtrsim 15$~pc in size 
and contains $\gtrsim 2 \times 10^5~M_\odot$ of hydrogen.
While it exhibits very high (positive and negative) turbulent velocities,
its bulk radial velocity typically ranges from $\approx 40~{\rm km~s}^{-1}$
to $\approx 70~{\rm km~s}^{-1}$. 
Along the line of sight, it lies immediately behind Sgr~A$^*$.
Clearly, the Sgr A East Core contains the gas-and-dust shell observed by 
\cite{mezger&zsw_89}, and it can be identified with the GMC inside which 
the supernova explosion that created Sgr~A East took place.\footnote{
The designation Sgr~A East Core was taken up by \cite{mezger&dz_96}.
However, instead of envisioning an extended M$-$0.02$-$0.07
that would contain the entire Sgr~A East Core (in addition to 
the Curved Streamer), they regarded M$-$0.02$-$0.07 as being only
"the compressed eastern part of the Sgr A East Core".
}

\noindent
$\bullet$ The Curved Streamer, which stretches east of Sgr~A East 
from the northern end of M$-$0.13$-$0.08 up to the eastern part of 
the Sgr~A East Core, is $\approx 7.5$~pc wide in $b$, 
contains $\approx (1-1.5) \times 10^5~M_\odot$ of hydrogen
and has radial velocities increasing steeply with Galactic longitude
from $\approx 25~{\rm km~s}^{-1}$ to $\approx 65~{\rm km~s}^{-1}$.
It lies in front of Sgr~A$^*$,
with its southern end connecting with M$-$0.13$-$0.08
and its northern end pointing deeper inward, though probably 
not deep enough to connect with the Sgr~A East Core.

\noindent
The above GMCs are embedded in a clumpy and highly turbulent 
molecular intercloud medium, 
with estimated hydrogen mass $\sim 10^6~M_\odot$ (in the inner $\sim 50$~pc),
average density $\sim 10^2~{\rm cm}^{-3}$ 
and radial velocities in the range $\approx -40$ to $+90~{\rm km~s}^{-1}$.

\cite{serabyn&la_92}, who mapped M$-$0.02$-$0.07 
in the CS $J = 5 \to 4$ and $J = 7 \to 6$ transitions
(at 245~GHz and 343~GHz, respectively),
also came to the conclusion that this cloud consists of two components:
a dense molecular core peaking at $(\Delta \alpha, \Delta \delta) 
\approx (3\fmn0,1\fmn5)$ with respect to Sgr~A$^*$ 
and a molecular ridge curving all the way around 
the eastern edge of the Sgr~A East radio shell.
As noted earlier by other authors, this spatial configuration
strongly suggests that Sgr~A East is colliding with, and compressing, 
M$-$0.02$-$0.07.
\cite{serabyn&la_92} found that radial velocities in M$-$0.02$-$0.07 
peak at $\approx 45~{\rm km~s}^{-1}$ and span the range 
$\approx (25 - 65)~{\rm km~s}^{-1}$,
with a steady increase from south to north along the ridge.
They argued that this velocity gradient is intrinsic to the cloud,
and not caused by its interaction with Sgr~A East.
From the measured CS $(7 \! \to \! 6)$-to-$(5 \! \to \! 4)$ line ratio, 
they derived a true H$_2$ density $\approx (1-2) \times 10^6~{\rm cm}^{-3}$,
and from the CS velocity-integrated line intensities, they derived
a space-averaged H$_2$ density $\approx 3 \times 10^4~{\rm cm}^{-3}$
near the emission peak and lower by up to a factor $\approx 2$ in the ridge
(assuming a line-of-sight depth $\approx 2.5$~pc)
as well as a hydrogen mass $\approx 1.5 \times 10^5~M_\odot$ 
for the entire cloud.

In contrast to \cite{zylka&mw_90}, who viewed the Curved Streamer
(their $^{13}$CO counterpart of the CS ridge) as a northward extension 
of M$-$0.13$-$0.08, separate from the Sgr~A East Core,
\cite{serabyn&la_92} concluded that the CS ridge is truly part 
of M$-$0.02$-$0.07 and in physical contact with Sgr~A East, 
while being separate from M$-$0.13$-$0.08.
Moreover, since they observed both highly blue- and redshifted gas 
just inside of the compressed CS ridge, which they identified with 
gas accelerated by the expansion of Sgr~A East, they concluded that 
molecular gas must be present on both the near and far sides of Sgr~A East 
(although with asymmetric distributions in favor of the far side).
In other words, the gas-and-dust shell surrounding Sgr~A East 
may be thinner \citep[as suggested by][]{mezger&zsw_89},
but not completely open, toward the Sun.

Yet another perspective emerges from the work of \cite{coil&h_99, coil&h_00},
who observed the central $10~{\rm pc} \times 15~{\rm pc}$ of the Galaxy 
in the NH$_3$ $(J,K) = (1,1)$ and $(2,2)$ transitions (both around 23.7~GHz).
Aside from an incomplete ring of emission corresponding to the CNR,
they clearly saw two long and narrow molecular streamers 
located to the south and east, respectively, of Sgr~A$^*$
and running roughly parallel to the Galactic plane.
\cite{coil&h_99} focused on the Southern Streamer and
\cite{coil&h_00} on the Eastern Streamer, which they also called
the Molecular Ridge.

The Southern Streamer, which is $\approx 10~{\rm pc} \times 2~{\rm pc}$
in projection, appears to connect the northern edge of 
the $20~{\rm km~s}^{-1}$ cloud to the southeastern part of the CNR
\citep{coil&h_99}.
Its bulk radial velocity is $\approx (20 - 35)~{\rm km~s}^{-1}$,
with a systematic increase toward the CNR,
while its velocity dispersion is $\approx (30 - 40)~{\rm km~s}^{-1}$,
with an abrupt jump to $\gtrsim 50~{\rm km~s}^{-1}$ near the CNR.
The gas kinetic temperature is $\approx (17 - 35)$~K in most of 
the streamer and jumps to $\approx 300$~K at its northern tip.
Taken together, these morphological, kinematic and thermal properties 
provide good evidence that the Southern Streamer is feeding the CNR
with molecular gas from the $20~{\rm km~s}^{-1}$ cloud
-- as suggested before by \cite{okumura&ifh_91}.
In this scenario, the northward velocity gradient measured along 
the Southern Streamer automatically positions the $20~{\rm km~s}^{-1}$ cloud 
in front of the CNR.
As seen in NH$_3$ emission, the Southern Streamer has
a hydrogen mass $\sim 3.5 \times 10^5~M_\odot$ 
and, assuming a line-of-sight depth $\approx 2$~pc, 
a space-averaged H$_2$ density $\sim (1-2) \times 10^5~{\rm cm}^{-3}$.
For comparison, the true H$_2$ densities traced by the NH$_3$
$(1,1)$ and $(2,2)$ transitions are typically a few $10^5~{\rm cm}^{-3}$.

The Eastern Streamer, or Molecular Ridge, which is somewhat longer 
($\gtrsim 12~{\rm pc}$) in projection than the Southern Streamer,
appears to trace the denser parts of the $50~{\rm km~s}^{-1}$ cloud
\citep{coil&h_00}.
Its northern half wraps around the eastern edge of Sgr~A East,
while its southern half continues past Sgr~A East 
toward the $20~{\rm km~s}^{-1}$ cloud.
Its bulk radial velocity globally jumps 
from $\approx 40~{\rm km~s}^{-1}$ in the northern half
to $\approx 20~{\rm km~s}^{-1}$ in the southern half,
with a general tendency to increase westward in the northern half
and eastward in the southern half. 
However, there are regions in the northern half which display 
both blue- and (brighter) redshifted emission,
suggesting that the Molecular Ridge contains gas both in front of
and (in greater amounts) behind Sgr~A East.
The velocity dispersion, too, has a discontinuous behavior:
it turns from roughly uniform in the northern half,
consistent with the gas being processed and postshock,
to highly variable in the southern half, consistent with the gas being 
strongly perturbed by the G359.92$-$0.09 SNR to the south of Sgr~A East.
This SNR appears to be also interacting with Sgr~A East, 
producing a noticeable inward bend in its southern boundary,
and with the $20~{\rm km~s}^{-1}$ cloud, making its eastern edge
sharp and straight.
Since the SNR is $\approx 3\fmn5$ (9~pc) in size, 
these simultaneous interactions imply that the $20~{\rm km~s}^{-1}$ cloud 
must be less than $\approx 9$~pc away from (i.e., in front of) Sgr~A East.
As seen in NH$_3$ emission, the Molecular Ridge has
a hydrogen mass $\sim 1.5 \times 10^5~M_\odot$, 
with $\sim 1.1 \times 10^5~M_\odot$ in the northern half
and $\sim 0.4 \times 10^5~M_\odot$ in the southern half,
and a space-averaged H$_2$ density $\sim (1-2) \times 10^5~{\rm cm}^{-3}$,
assuming again a line-of-sight depth $\approx 2$~pc.

A follow-up study of the central 10~pc of the Galaxy was performed 
by \cite{mcgary&co_01}, based on spectral observations
of the NH$_3$ $(1,1)$, $(2,2)$ and $(3,3)$ transitions 
(all between 23.7~GHz and 23.9~GHz).
In addition to the Southern Streamer and the Molecular Ridge 
clearly visible in the $(1,1)$ and $(2,2)$ maps of
\cite{coil&h_99, coil&h_00}, the $(3,3)$ map brings out two new features:

\noindent
$\bullet$ The Western Streamer, located west of Sgr~A$^*$,
extends in the north-south direction over $\approx 2\fmn8$ (7~pc).
It closely follows the western boundary of Sgr~A East,
suggesting that it is made of material swept up by the expansion of the SNR.
Its bulk radial velocity gradually increases
from $\approx -70~{\rm km~s}^{-1}$ near its southern tip
to $\approx +90~{\rm km~s}^{-1}$ near its northern tip.
This large velocity gradient could be due to intrinsic rotation 
or to orbital motion around Sgr~A$^*$, with a possible enhancement 
by the expansion of Sgr~A East.

\noindent
$\bullet$ The Northern Ridge, located northeast of Sgr~A$^*$,
extends in the northeast-southwest direction over $\approx 1\fmn4$ (3.5~pc).
It lies along the northern boundary of Sgr~A East,
so it, too, could be made of swept-up material.
Its bulk radial velocity is $\approx -10~{\rm km~s}^{-1}$ all along,
and it is kinematically connected to the northeastern lobe of the CNR 
through a narrow streamer along which the radial velocity increases 
smoothly from $\approx -10~{\rm km~s}^{-1}$ 
to $\approx +60~{\rm km~s}^{-1}$.
If this kinematic connection represents inflow toward the CNR,
the Northern Ridge must be slightly in front of the northeastern lobe 
of the CNR, and Sgr~A East itself must be close to the CNR.

\cite{herrnstein&h_05} pursued \citeauthor{mcgary&co_01}'s 
(\citeyear{mcgary&co_01}) study
after adding to their NH$_3$ $(1,1)$, $(2,2)$ and $(3,3)$ data
the 25~GHz NH$_3$ $(6,6)$ data of \cite{herrnstein&h_02}.
Assuming an NH$_3$-to-H$_2$ ratio of $10^{-7}$
\citep[as opposed to $10^{-8}$ in][]{coil&h_99, coil&h_00},
they estimated the H$_2$ masses of the major molecular clouds at roughly 
$\gtrsim 8 \times 10^4~M_\odot$ for the Southern Streamer,
$\gtrsim 3 \times 10^4~M_\odot$ for the Molecular Ridge,
$\gtrsim 5 \times 10^4~M_\odot$ for the core of the $50~{\rm km~s}^{-1}$ cloud,
$\sim 4 \times 10^3~M_\odot$ for the Western Streamer
and $\sim 2 \times 10^3~M_\odot$ for the Northern Ridge.
Note that the first three values only give lower limits 
because the associated objects extend past the edge of the NH$_3$ maps
-- this is particularly true of the core of the $50~{\rm km~s}^{-1}$ cloud,
which has roughly three-quarters of its projected volume outside the maps,
so that its total mass could actually be as large as 
$\sim 2 \times 10^5~M_\odot$.
\cite{herrnstein&h_05} also found the molecular gas to have 
a two-temperature structure on scales $\lesssim 0.5$~pc, 
with $\sim 75\%$ of the gas at $\lesssim 25$~K (probably $\sim 15$~K) 
and $\sim 25\%$ at $\sim 200$~K.

By considering their NH$_3$ data in conjunction with existing data
at other frequencies, \cite{herrnstein&h_05} developed
a three-dimensional picture for the spatial arrangement of the main 
molecular features within a few pc of the GC.
In this picture, Sgr~A$^*$, Sgr~A West and the surrounding CNR 
reside just inside the near surface of Sgr A East;
the $20~{\rm km~s}^{-1}$ cloud and the Southern Streamer lie
entirely in front of Sgr~A East;
the $50~{\rm km~s}^{-1}$ cloud envelops Sgr~A East from front to back 
along its eastern side, and it is connected to the $20~{\rm km~s}^{-1}$ 
cloud by the Molecular Ridge.
These line-of-sight positions are consistent with the observational 
results presented above as well as with the finding that 
the $20~{\rm km~s}^{-1}$ cloud strongly absorbs the $(2-10)$~keV X-ray 
emission from the central $17'$, while the $50~{\rm km~s}^{-1}$ cloud 
does not \citep{park&mbm_04}.
Furthermore, from the $\approx -70$ to $+90~{\rm km~s}^{-1}$
velocity gradient along the Western Streamer
and the $\approx -10~{\rm km~s}^{-1}$ velocity all along the Northern Ridge
measured by \cite{mcgary&co_01}, \cite{herrnstein&h_05} concluded 
that the Western Streamer is highly inclined to the plane of the sky, 
with its southern part on the front side and its northern part 
on the back side of Sgr~A East,\footnote{\label{foot}
This interpretation raises a self-consistency problem. 
If the Western Streamer is indeed swept-up material at the surface of 
Sgr~A East, the mere fact that it runs precisely along the projected
boundary of Sgr~A East provides strong evidence that 
it is approximately contained in the sky plane through the center of Sgr~A East.
This line-of-sight position is supported by the conclusions of 
\cite{lee&pcd_08} (see below in Section~\ref{obs_clouds}).
}
while the Northern Ridge is roughly parallel to the plane of the sky,
at about the line-of-sight distance of Sgr~A East's center.

The three-dimensional picture of \cite{herrnstein&h_05} was later 
modified by \cite{lee&pcd_08}, who carried out spectral 
observations of the $2.12~\mu$m H$_2$ $v=1-0 \ S(1)$ emission line
in four areas along the periphery of Sgr~A East,
where the shock front is expanding in molecular clouds.
In contrast to NH$_3$ emission, which traces cool ($T \lesssim 100$~K)
molecular gas, H$_2$ emission traces hot ($T \sim 2000$~K) molecular gas
and is, therefore, ideally suited to probe shock-heated molecular regions.
Thus \cite{lee&pcd_08} compared position-velocity diagrams 
of H$_2$ emission (assumed to trace postshock gas) 
and NH$_3$ emission (assumed to trace preshock gas)
to derive shock velocities and use them as indicators 
of line-of-sight positions relative to the center of Sgr~A East.
As a general rule, they found that the H$_2$ lines are much broader 
and either blue- or redshifted with respect to the corresponding NH$_3$ lines.
They concluded that Sgr~A East is driving shocks into,
and hence is in physical contact with, each of the $50~{\rm km~s}^{-1}$ cloud,
the Molecular Ridge (at least its northern part), the Northern Ridge, 
the Western Streamer, the Southern Streamer and the CNR. 
More specifically, they gathered that 
the $50~{\rm km~s}^{-1}$ cloud brackets the eastern part of Sgr~A East
along the line of sight; 
the Molecular Ridge probably lies between the front and back sides
of Sgr~A East, with its northern end slightly tipped toward the back;
the Northern Ridge lies to the back side of Sgr~A East;
the northern half of the Western Streamer surrounds 
the western edge of Sgr~A East from front to back;
the Southern Streamer and the CNR lie in front of Sgr~A East.
They also confirmed earlier claims that the Molecular Ridge connects
the $20~{\rm km~s}^{-1}$ and $50~{\rm km~s}^{-1}$ clouds,
while the Southern Streamer connects the $20~{\rm km~s}^{-1}$ cloud 
to the CNR.

\cite{amo&mm_11} obtained complementary information 
on the three-dimensional disposition of molecular features 
in the central 12~pc and on their possible connections
by comparing the emissions from selected molecular tracers
believed to respond differently to interstellar shocks 
and to UV radiation.
Their data set comprises their own measurements of 
SiO $J = 2 \to 1$ (tracer of shocked gas), HNCO $J = 5_{0.5} \to 4_{0.4}$ 
(tracer of shocked gas, very sensitive to photo-dissociation),
H$^{13}$CO$^+$ $J = 1 \to 0$ (similar to SiO)
and HN$^{13}$C $J = 1 \to 0$ (similar to HNCO),
in addition to \citeauthor{tsuboi&hu_99}'s (\citeyear{tsuboi&hu_99})
measurements of CS $J = 1 \to 0$ (tracer of quiescent dense gas).
They used the HNCO-to-SiO, SiO-to-CS and HNCO-to-CS intensity ratios
as indicators of relative distances to the central star cluster
(the source of the UV radiation responsible for photo-dissociation)
and of the presence of gas shocked by the expansion of Sgr~A East.
In this manner, they found that
the Molecular Ridge is probably relatively distant from the central cluster 
or else shielded from its UV photons; 
the Northern Ridge is close to the central cluster
and possibly connected to the northeastern part of the CNR;
the Southern Streamer approaches the central cluster going northward 
and probably connects the $20~{\rm km~s}^{-1}$ cloud to the southeastern 
part of the CNR;
the Western Streamer is close to the central cluster and was swept up
by the Sgr~A East shock.

It is important to emphasize that the relative line-of-sight locations
of the main interstellar objects near the GC are still a matter of
controversy.
For instance, the 18~cm spectral observations of the four OH ground-state
transitions by \cite{karlsson&ssw_03} showed strong absorption against 
the eastern and most of the western parts of the Sgr~A East shell, 
but a lack of absorption against the spiral pattern of the Sgr~A West 
H\,{\sc ii} region.
This prompted the authors to suggest that a fraction of the molecular belt
(comprising the $20~{\rm km~s}^{-1}$ and $50~{\rm km~s}^{-1}$ clouds)
lies in front of Sgr~A East and, at the same time, behind Sgr~A West, 
so that both radio sources must be separated by a finite distance
along the line of sight.
Obviously, this view contradicts the notion that Sgr~A West
is embedded within Sgr~A East \citep[e.g.,][]{yusef&mw_00, maeda&bfm_02}.
Similarly, the 1720~MHz OH absorption measurements of \cite{sjouwerman&p_08}
indicated that the $20~{\rm km~s}^{-1}$ and $50~{\rm km~s}^{-1}$ clouds 
lie mostly behind Sgr~A West and at least partly in front of Sgr~A East.
Incidentally, both \cite{karlsson&ssw_03} and \cite{sjouwerman&p_08}
were able to clearly identify the CNR in OH absorption at high absolute
velocities, thereby confirming its location on the near side of Sgr~A East.

\section{\label{model_gas}Our representation of the interstellar gas}

Now armed with all the observational results described in
Section~\ref{obs_gas},
we proceed to construct a plausible and handy (as far as possible)
representation of the interstellar gas within $\sim 10$~pc of Sgr~A$^*$.
Unless stated otherwise, the parameter values adopted in the following
subsections are based on the observational studies discussed in 
the corresponding subsections of Section~\ref{obs_gas}
and on complementary theoretical arguments made both to fill in
the gaps in the observational estimates and to ensure self-consistency 
of our gas representation.
For convenience, a summary of our adopted values for the geometrical 
and thermodynamic parameters of the different structural components 
is given in Tables~\ref{summary_geo} and \ref{summary_thermo}.
In addition, three orthogonal (front, side and top) views
showing the spatial organization of the different components, 
with their respective shapes and relative sizes, are schematically 
drawn in Figure~\ref{geometry}.

\begin{table*}[t]
\caption{Geometrical parameters
of the different structural components in our representation of 
the interstellar gas.
}
\label{summary_geo}
\begin{minipage}[t]{2\columnwidth}
\centering
\renewcommand{\footnoterule}{}  
\begin{tabular}{lllcl}
\hline
\hline
\noalign{\smallskip}
Component & Shape & 
Dimensions [pc]\footnote{
The errors in our adopted dimensions, which arise from observational 
uncertainties and from our geometrical approximations,
are estimated as follows [in pc]:
Central Cavity: $\delta l_y \approx \delta l_z \approx 0.5$.
Minispiral: $\delta l \approx 4$, $\delta d \approx \, _{-0.05}^{+0.1}$ .
Neutral streamers: $\delta l \approx 3$, $\delta d \approx 0.15$.
Central sphere: $\delta d \approx 0.4$.
CNR: $\delta r_{\rm in} \approx 0.4$,
$\delta r_{\rm out} \approx \, _{-0.5}^{+4}$ ,
$\delta h_{\rm in} \approx 0.2$, $\delta h_{\rm out} \approx 0.5$.
Sgr~A East: $\delta L_y \approx \delta L_z \approx 1.5$.
Radio halo: $\delta d \approx 5$.
M$-$0.13$-$0.08: $\delta \ell_{y'} \approx \delta \ell_{z'} \approx 3$.
M$-$0.02$-$0.07: $\delta d \approx 3$.
Swept-up shell: $\delta \Delta r \approx 0.7$. 
Bridge and Southern Streamer: $\delta d \approx 1$.
Western Streamer and Northern Ridge: $\delta l \approx 2$, 
$\delta d \approx 0.5$.
} & 
Volume [pc$^3$] & 
Position [pc]\footnote{
The errors in our adopted positions are estimated as follows [in pc]:
$\approx 0.1$ for the Central Cavity;
$\approx 0.3$ for the CNR and Sgr~A East;
$\sim 2$ for the radio halo and the different molecular clouds in the belt.
}\\
\noalign{\smallskip}
\hline
\noalign{\smallskip}
Central Cavity \ (CC) & ellipsoid &
$l_x \times l_y \times l_z = 2.9 \times 2.9 \times 2.1$ & 
9.2 & 
centered on Sgr~A$^*$ \\
\quad Extended component & & & 
8.4 & 
 \\
\quad Minispiral & & 
$l = 9.7$ \ \& \ $d = 0.1$ &
0.076 &   
 \\
\quad Neutral streamers & & 
$l = 6.2$ \ \& \ $d = 0.3$ &
0.44 &   
 \\
\quad Central sphere & & 
$d = 0.8$ &
0.27 &   
 \\
\noalign{\smallskip}
\hline
\noalign{\smallskip}
Circumnuclear Ring \ (CNR) & trapezoidal ring & 
$r_{\rm in} = 1.2$, \ $r_{\rm out} = 3.0$ \ \& \ 
$h_{\rm in} = 0.4$, \ $h_{\rm out} = 1.0$ \hspace*{-0.4cm} &
18 & 
centered on Sgr~A$^*$ \\
\quad Main molecular ring & & 
$r_{\rm in} = 1.2$, \ $r_{\rm out} = 3.0$ &
18 &   
 \\
\quad Photo-dissociated inner layer \hspace*{-0.2cm} & & 
$r_{\rm in} = 1.2$, \ $r_{\rm out,a} = 1.6$ &
1.65 &   
 \\
\noalign{\smallskip}
\hline
\noalign{\smallskip}
Sgr~A East SNR & ellipsoid & 
$L_x \times L_y \times L_z = 9.0 \times 9.0 \times 6.7$ &
285 & 
$(x_{\rm c}, y_{\rm c}, z_{\rm c}) = (-2.0, 1.2, -1.5)$ \\
\quad Extended component & & &
260 &   
 \\
\noalign{\smallskip}
\hline
\noalign{\smallskip}
Radio halo & sphere &
$d = 18$ &
3\,050 & 
$(x_{\rm c}, y_{\rm c}, z_{\rm c}) = (-2.0, 1.2, -1.5)$ \\
\quad Extended component & & &
2\,440 &   
 \\
\noalign{\smallskip}
\hline
\noalign{\smallskip}
Belt of molecular clouds\footnote{
The swept-up shell is indicated twice, because it is part of both
M$-$0.02$-$0.07 and the Molecular Ridge.
}\footnote{
The ranges given for the lengths and volumes of the Bridge
(and hence the Molecular Ridge) and the Southern Streamer 
correspond to the range in the line-of-sight coordinate of the center 
of M$-$0.13$-$0.08, $x_{\rm \scriptscriptstyle SC}$.
} & & & & \\
\quad M$-$0.13$-$0.08 \ (SC) & ellipsoid &
$\ell_{x} \times \ell_{y'} \times \ell_{z'} = 7.5 \times 15 \times 7.5$ &
442 &  
$(x_{\rm \scriptscriptstyle SC}, y_{\rm \scriptscriptstyle SC}, 
z_{\rm \scriptscriptstyle SC}) = (4-12, -11, -5)$ \\
\quad M$-$0.02$-$0.07 \ (EC) & indented sphere &
$d = 9$ &
356 & 
$(x_{\rm \scriptscriptstyle EC}, y_{\rm \scriptscriptstyle EC}, 
z_{\rm \scriptscriptstyle EC}) = (-3, 7, -4.5)$ \\
\qquad Preshock core & & &
305 & 
 \\
\qquad Swept-up shell & &
$\Delta r = 1.5$ & 
51 &  
along NE bdy of SNR \\
\quad Molecular Ridge \ (MR) & & & 
$79-98$  
 \\
\qquad Swept-up shell & &
$\Delta r = 1.5$ & 
51 &  
along NE bdy of SNR \\
\qquad Bridge & curved cylinder &
$l = (9-15)$ \ \& \ $d = 2$ &
$28-47$ & 
between EC \& SC \\
\quad Southern Streamer \ (SS) & curved cylinder &
$l = (7-14)$ \ \& \ $d = 2$ &
$22-44$ & 
between SC \& CNR \\
\quad Western Streamer \ (WS) & curved cylinder &
$l = 8$ \ \& \ $d = 1$ &
6 & 
along W bdy of SNR \\
\quad Northern Ridge \ (NR) & curved cylinder &
$l = 4$ \ \& \ $d = 1$ &
3 & 
along N bdy of SNR \\
\noalign{\smallskip}
\hline
\end{tabular}
\end{minipage}
\end{table*}

\begin{table*}[t]
\caption{Thermodynamic parameters (interstellar phase, temperature,
mean hydrogen density and hydrogen mass)
of the different structural components in our representation of 
the interstellar gas.
}
\label{summary_thermo}
\begin{minipage}[t]{2\columnwidth}
\centering
\renewcommand{\footnoterule}{}  
\begin{tabular}{llccc}
\hline
\hline
\noalign{\smallskip}
Component & Phase & 
$T~[K]$\footnote{
The actual temperature ranges, accounting for observational uncertainties,
model approximations and true physical dispersion, are [in K]:
Central Cavity: $\approx (5\,000 - 13\,000)$ for the warm ionized gas;
$\approx (100-240)$ for the atomic gas;
$\approx (1.4-2.3) \times 10^7$ for the hot ionized gas.
CNR: $\approx (40-300)$ for the molecular gas;
$\approx (200 - 1\,000)$ for the atomic gas.
Sgr~A East: $\approx (1-7) \times 10^7$.
Radio halo: $\approx (5\,000 - 10\,000)$.
Belt of molecular clouds: $\approx (15-200)$.
} & $n_{\rm H}~[{\rm cm}^{-3}]$\footnote{
The actual hydrogen density ranges, accounting for observational uncertainties,
model approximations and true physical dispersion, are [in cm$^{-3}$]:
Central Cavity: $\approx (10^3 - 2 \times 10^5)$ for the warm ionized gas;
$\approx (2 \times 10^3 - 3 \times 10^5)$ for the atomic gas;
$\approx (10-40)$ for the hot ionized gas.
CNR: $\approx (10^4 - 2 \times 10^8)$ for the molecular gas;
$\approx (10^4 - 2 \times 10^5)$ for the atomic gas.
Sgr~A East: $\approx (1.5-30)$.
Radio halo: $\approx (10 - 1\,000)$.
Belt of molecular clouds: $\approx (10^4 - 4 \times 10^6)$.
} & $M_{\rm H}~[M_\odot]$\footnote{
The errors in our adopted masses are estimated as follows:
a factor $\sim 4$ for the ionized components 
(including uncertainties in the observed emission measures, 
in the projected surface areas and in the line-of-sight depths);
a factor $\sim 5$ for the atomic and molecular components 
(including uncertainties in the measured column densities, 
in the tracer-to-hydrogen ratios and in the projected surface areas).
For the radio halo, the error is in fact larger than a factor $\sim 4$,
as the relative contribution from warm ionized gas to the observed 
radio emission is itself very uncertain.
} \\
\noalign{\smallskip}
\hline
\noalign{\smallskip}
Central Cavity \ (CC) & & & & \\
\quad Extended component & warm ionized & 
7\,000 & 910 & 190 \\
\quad Minispiral & warm ionized & 
7\,000 & 6\,200 & 12 \\
\quad Neutral streamers & atomic & 
170 & $1.5 \times 10^4$ & 160 \\
\quad Central sphere & hot ionized & 
$1.5 \times 10^7$ & 18.5 & 0.12 \\
\noalign{\smallskip}
\hline
\noalign{\smallskip}
Circumnuclear Ring \ (CNR) & & & & \\
\quad Main molecular ring & molecular & 
150 & $4.4 \times 10^5$ & $2 \times 10^5$ \\
\quad Photo-dissociated inner layer & atomic & 
300 & $3.2 \times 10^4$ & 1\,300 \\
\noalign{\smallskip}
\hline
\noalign{\smallskip}
Sgr~A East SNR & & & & \\
\quad Extended component & hot ionized &
$1.5 \times 10^7$ & 3.0 & 19 \\
\noalign{\smallskip}
\hline
\noalign{\smallskip}
Radio halo & & & & \\
\quad Extended component & warm ionized &
7\,000 & 210 & $1.3 \times 10^4$ \\
\noalign{\smallskip}
\hline
\noalign{\smallskip}
Belt of molecular clouds\footnote{
The swept-up shell is indicated twice, because it is part of both
M$-$0.02$-$0.07 and the Molecular Ridge.
}\footnote{
The mass ranges given for the Bridge (and hence the Molecular Ridge)
and the Southern Streamer correspond to the range in
the line-of-sight coordinate of the center of M$-$0.13$-$0.08,
$x_{\rm \scriptscriptstyle SC} = (4-12)$~pc.
} & & & & \\
\quad M$-$0.13$-$0.08 \ (SC) & molecular &
60 & $2 \times 10^4$ & $2.2 \times 10^5$ \\
\quad M$-$0.02$-$0.07 \ (EC) & molecular &
60 & & $1.9 \times 10^5$ \\
\qquad Preshock core & &
& $2 \times 10^4$ & $1.5 \times 10^5$ \\
\qquad Swept-up shell & &
& $3 \times 10^4$ & $3.8 \times 10^4$ \\
\quad Molecular Ridge \ (MR) & molecular & 
60 & $3 \times 10^4$ & $(5.9-7.3) \times 10^4$ \\
\qquad Swept-up shell & &
& & $3.8 \times 10^4$ \\
\qquad Bridge & &
& & $(2.1-3.5) \times 10^4$ \\
\quad Southern Streamer \ (SS) & molecular &
60 & $3 \times 10^4$ & $(1.6-3.2) \times 10^4$ \\
\quad Western Streamer \ (WS) & molecular &
60 & $3 \times 10^4$ & $4.5 \times 10^3$ \\
\quad Northern Ridge \ (NR) & molecular &
60 & $3 \times 10^4$ & $2.2 \times 10^3$ \\
\noalign{\smallskip}
\hline
\end{tabular}
\end{minipage}
\end{table*}

\subsection{\label{model_cavity}The Central Cavity}

We approximate the Central Cavity as an ellipsoid centered on Sgr~A$^*$,
axisymmetric about the vertical axis and having dimensions 
$l_x \times l_y \times l_z
= 2.9~{\rm pc} \times 2.9~{\rm pc} \times 2.1~{\rm pc}$,
where $l_y$ and $l_z$ correspond to the projected 
{\it FWHM} dimensions of the extended radio component discussed 
by \cite{beckert&dmz_96}.
The volume of the Central Cavity is then 
$V_{\rm \scriptscriptstyle CC} = 9.2~{\rm pc}^3$.
We consider that the Central Cavity contains warm ionized gas,
divided between an extended (or diffuse) component and the Minispiral, 
neutral atomic gas, confined to one or two neutral streamers,
and hot ionized gas, extending over the central 0.8~pc.
For simplicity, we ignore the fine-scale structure observed in each 
of these gas components (including the Minicavity, stellar winds, 
dense clumps, etc.), which we take to be smooth and homogeneous.
We set the temperatures of the three gases to
$T_{\rm wi} = 7\,000$~K \citep{roberts&g_93, shukla&ys_04},
$T_{\rm a} = 170$~K \citep{jackson&ggr_93}
and $T_{\rm h} = 1.5 \times 10^7$~K \citep{baganoff&mmb_03},
and we consider that the warm ionized gas has hydrogen completely ionized 
and helium completely neutral, while the hot ionized gas has both
hydrogen and helium fully ionized.

The extended component of the warm ionized gas is supposed to have
a central emission measure of $2.2 \times 10^6~{\rm pc~cm}^{-6}$
(value derived by \cite{beckert&dmz_96} and scaled up to our adopted 
temperature), which, combined with a line-of-sight dimension
$l_x = 2.9~{\rm pc}$ and a filling factor $\phi_{\rm ext}$,
yields a true electron density $(n_{\rm e})_{\rm ext} = 
(870~{\rm cm}^{-3}) \, \phi_{\rm ext}^{-1/2}$
and a space-averaged electron density $(\langle n_{\rm e} \rangle)_{\rm ext} 
= (870~{\rm cm}^{-3}) \, \phi_{\rm ext}^{1/2}$.
Since all the free electrons are presumed to come from hydrogen,
the H$^+$ density is simply $n_{\rm H^+} = n_{\rm e}$
and the H$^+$ mass of the extended component 
$(M_{\rm H^+})_{\rm ext} = (200~M_\odot) \, \phi_{\rm ext}^{1/2}$.
The reason why our electron densities and H$^+$ mass differ somewhat
from those derived by \cite{beckert&dmz_96} is not only because 
we adopted a higher temperature, but also, and more importantly,
because we assumed an ellipsoidal cavity with a central line-of-sight 
depth of 2.9~pc, whereas they assumed a constant line-of-sight depth 
$\approx 1$~pc.

The parameters of the Minispiral are more uncertain.
Here, we consider that the Minispiral is composed of three arms:
the Northern Arm, the Eastern Arm + Bar and the Western Arc.
We furthermore assume that the three arms follow the Keplerian elliptical orbits
derived by \cite{zhao&mga_09} -- in a particularly careful analysis 
based on a combination of proper motion and radial velocity measurements --
and rescaled to $r_\odot = 8.5$~kpc
(see their Table~5 for a list of all the orbital parameters),
that they are, respectively, 2.7~pc, 3.5~pc and 3.5~pc long
\citep[as calculated from the ranges of true anomaly 
quoted by][]{zhao&mga_09}
and that they all have a circular cross-section of diameter
$(0.1~{\rm pc}) \, d_{0.1}$.
Under these conditions, the total length of the Minispiral is 9.7~pc
and its total volume 
$V_{\rm msp} = (0.076~{\rm pc}^3) \, d_{0.1}^2$.
We note that the total length of 9.7~pc obtained here
is significantly greater than the total length of 3.9~pc 
measured by \cite{shukla&ys_04},
which is mainly because the latter is a two-dimensional length 
in the plane of the sky, as opposed to a full three-dimensional length.

Once the volume of the Minispiral has been determined,
its electron density and H$^+$ mass can be inferred from 
its measured radio emission flux density.
Indeed, for a warm ionized gas component (with given electron temperature 
and ionization state) occupying a volume $V$, 
within which its filling factor is $\phi$,
and producing thermal free-free emission with flux density $F_\nu$, 
the true electron density and H$^+$ mass scale roughly as
$n_{\rm e} \propto \left[ F_\nu / (V \, \phi) \right]^{1/2}$ and 
$M_{\rm H^+} \propto \left[ F_\nu \, (V \, \phi) \right]^{1/2}$.
\cite{beckert&dmz_96}, who tried to separate the Minispiral 
from the extended component in a 2~cm continuum map of Sgr~A West,
estimated their contributions to the 2~cm flux density
at $(F_\nu)_{\rm ext} = 19$~Jy and $(F_\nu)_{\rm msp} = 8$~Jy.
For the effective volumes, we have
$(V \, \phi)_{\rm ext} 
= (9.2~{\rm pc}^3) \, \phi_{\rm ext}$
and $(V \, \phi)_{\rm msp} 
= (0.076~{\rm pc}^3) \, d_{0.1}^2 \, \phi_{\rm msp}$.
We also know that the extended component has
$(n_{\rm e})_{\rm ext} = 
(870~{\rm cm}^{-3}) \, \phi_{\rm ext}^{-1/2}$
and $(M_{\rm H^+})_{\rm ext} = 
(200~M_\odot) \, \phi_{\rm ext}^{1/2}$.
It then follows that the Minispiral must have 
$(n_{\rm e})_{\rm msp} = 
(6\,200~{\rm cm}^{-3}) \, d_{0.1}^{-1} \, \phi_{\rm msp}^{-1/2}$ 
and $(M_{\rm H^+})_{\rm msp} = 
(12~M_\odot) \, d_{0.1} \, \phi_{\rm msp}^{1/2}$.
For $d_{0.1} = 1$ \citep{beckert&dmz_96, shukla&ys_04} 
and $\phi_{\rm msp} = 1$, 
the volume, (true or space-averaged) electron density and H$^+$ mass 
of the Minispiral reduce to $V_{\rm msp} = 0.076~{\rm pc}^3$,
$(n_{\rm e})_{\rm msp} = 6\,200~{\rm cm}^{-3}$
and $(M_{\rm H^+})_{\rm msp} = 12~M_\odot$.\footnote{
These values refer to the ionized gas traced by the 2~cm continuum emission.
It is, therefore, not surprising that the density derived here is lower
than the density inferred from emission at millimeter wavelengths 
\citep{shukla&ys_04, zhao&bmd_10},
which refers to a denser ionized gas component.
}
The density and mass are reasonably close to those obtained by 
\cite{beckert&dmz_96}, and the differences can mostly be explained 
by our using the full three-dimensional length of the Minispiral, 
instead of its two-dimensional length in the plane of the sky.
If we now split the H$^+$ mass of the Minispiral into its three arms 
according to their respective lengths, 
we find $3.4~M_\odot$ in the Northern Arm, 
$4.3~M_\odot$ in the Eastern Arm + Bar
and $4.3~M_\odot$ in the Western Arc.

Neutral atomic gas inside the Central Cavity resides either in one large
neutral streamer, the so-called Northern Streamer, bounded by the ionized 
Northern and Eastern Arms of the Minispiral \citep{jackson&ggr_93},
or in two thinner neutral streamers, adjacent to the Northern and 
Eastern Arms, respectively \citep{latvakoski&sgh_99}.
Because the Northern and Eastern Arms turn out to be on nearly
perpendicular orbits \citep{zhao&bmd_10}, it is hard to imagine 
that they could be linked to the same neutral streamer,
which leads us to opt for the second possibility and follow 
\cite{latvakoski&sgh_99}.
Nonetheless, to ensure self-consistency in our gas representation,
we do not strictly stick to their dust model, which places 
the neutral streamers on parabolic orbits about Sgr~A$^*$.
Instead, we consider that the neutral streamers run exactly alongside 
their ionized counterparts (themselves on Keplerian elliptical orbits; 
see above), on their sides farther from Sgr~A$^*$.
The neutral streamer associated with the Northern Arm is then 2.7~pc long 
and that associated with the Eastern Arm + Bar 3.5~pc long.
Next, we assume that both neutral streamers have a circular cross-section
of diameter 0.3~pc, which corresponds to their projected thickness 
in the far-infrared maps of \cite{latvakoski&sgh_99},
and we assign them the H\,{\sc i} masses derived by \cite{latvakoski&sgh_99}.
Thus we find that the neutral streamer associated with the Northern Arm
has a volume of 0.19~pc$^3$, an H\,{\sc i} mass of $110~M_\odot$,
and hence a mean H\,{\sc i} density of $2.3 \times 10^4~{\rm cm}^{-3}$,
while the neutral streamer associated with the Eastern Arm + Bar
has a volume of 0.25~pc$^3$, an H\,{\sc i} mass of $50~M_\odot$, 
and hence a mean H\,{\sc i} density of $8 \times 10^3~{\rm cm}^{-3}$.
Together, the two neutral streamers occupy a volume 
$V_{\rm nstr} = 0.44~{\rm pc}^3$, enclose an H\,{\sc i} mass 
$(M_{\rm H{\scriptscriptstyle I}})_{\rm nstr} = 160~M_\odot$
and have a mean H\,{\sc i} density
$(\langle n_{\rm H{\scriptscriptstyle I}} \rangle)_{\rm nstr} 
= 1.5 \times 10^4~{\rm cm}^{-3}$.
It is important to realize that the above mean H\,{\sc i} densities 
represent spatial averages over the neutral streamers,
whereas the space-averaged H\,{\sc i} density 
$\approx 1.6 \times 10^3~{\rm cm}^{-3}$ derived by \cite{jackson&ggr_93} 
refers to a spatial average over the Central Cavity.
Furthermore, if the true H\,{\sc i} density 
$\approx 3 \times 10^5~{\rm cm}^{-3}$
derived by \cite{jackson&ggr_93} is characteristic, 
the implied mean filling factor of [O\,{\sc i}]-emitting atomic gas 
within the neutral streamers is $\sim 0.05$.

Finally, for the hot ionized gas, we follow \cite{baganoff&mmb_03}
and assume that this gas is contained within a centered sphere 
of radius 0.4~pc, and hence volume $V_{\rm hot} = 0.27~{\rm pc}^3$,
that it is fully ionized, with twice solar abundances, 
and that it has a true electron density
$(n_{\rm e})_{\rm h} = (26~{\rm cm}^{-3}) \, \phi_{\rm h}^{-1/2}$.
Under these conditions, its true H$^+$ density is
$(n_{\rm H^+})_{\rm h} = \frac{1}{1.4} \, (n_{\rm e})_{\rm h} 
= (18.5~{\rm cm}^{-3}) \, \phi_{\rm h}^{-1/2}$
and its H$^+$ mass 
$(M_{\rm H^+})_{\rm hot} = (0.12~M_\odot) \, \phi_{\rm h}^{1/2}$.
To determine the filling factor $\phi_{\rm h}$, 
we consider the assumption of rough thermal pressure balance 
between the extended warm ionized gas and the hot ionized gas, such that
$2.2 \, (n_{\rm H^+})_{\rm ext} \, T_{\rm wi} \approx
2.6 \, (n_{\rm H^+})_{\rm h} \, T_{\rm h}$.
With $T_{\rm wi} = 7\,000$~K, $T_{\rm h} = 1.5 \times 10^7$~K
(see beginning of Section~\ref{model_cavity}),
$(n_{\rm H^+})_{\rm ext} = 
(870~{\rm cm}^{-3}) \, \phi_{\rm ext}^{-1/2}$ and
$(n_{\rm H^+})_{\rm h} = (18.5~{\rm cm}^{-3}) \, \phi_{\rm h}^{-1/2}$,
thermal pressure balance would imply $\phi_{\rm h} \gg \phi_{\rm ext}$,
which is impossible.\footnote{
Remember that $\phi_{\rm ext}$ is the filling factor of extended 
warm ionized gas within the Central Cavity, 
whereas $\phi_{\rm h}$ is the filling factor of hot ionized gas
inside the central 0.4~pc radius sphere alone. 
These filling factors must necessarily satisfy $\phi_{\rm ext} \ge 0.91$ 
(value obtained if all the extended warm ionized gas is excluded from 
the central sphere; see next paragraph) and $\phi_{\rm h} \le 1$, 
respectively, thereby ruling out $\phi_{\rm h} \gg \phi_{\rm ext}$.
}
In consequence, both gases cannot be in thermal pressure balance,
and the overpressured hot ionized gas will tend to completely fill
the central 0.4~pc radius sphere.
In other words, we may take $\phi_{\rm h} = 1$ inside the central sphere, 
whereupon the (true or space-averaged) H$^+$ density and H$^+$ mass 
of hot ionized gas become 
$(n_{\rm H^+})_{\rm h} = 18.5~{\rm cm}^{-3}$
and $(M_{\rm H^+})_{\rm hot} = 0.12~M_\odot$.

We can now complete our specification of the parameters of the extended 
warm ionized gas.
The volume left to this component is the volume inside the Central Cavity 
that is not occupied by either the Minispiral or the neutral streamers 
or the central sphere, i.e.,
$V_{\rm ext} = V_{\rm \scriptscriptstyle CC} - V_{\rm msp}
- V_{\rm nstr} - V_{\rm hot} = 8.4~{\rm pc}^3$,
and the associated filling factor within the Central Cavity is 
$\phi_{\rm ext} = V_{\rm ext} / V_{\rm \scriptscriptstyle CC} = 0.91$.
It then follows that the extended warm ionized gas has
a true H$^+$ density $(n_{\rm H^+})_{\rm ext} = 910~{\rm cm}^{-3}$,
a space-averaged H$^+$ density 
$(\langle n_{\rm H^+} \rangle)_{\rm ext} = 830~{\rm cm}^{-3}$
and an H$^+$ mass $(M_{\rm H^+})_{\rm ext} = 190~M_\odot$.

\subsection{\label{model_CNR}The Circumnuclear Ring}

Although all observations point to a highly irregular 
and clumpy structure, which extends significantly farther out 
to the southwest than to the northeast,
for simplicity we model the CNR as a well-defined, smooth 
and axisymmetric ring.
We consider that this ring is centered on Sgr~A$^*$
\citep[thus ignoring the slight southeast offset mentioned 
by][]{gusten&gwj_87}
and that it extends radially from $r_{\rm in} = 1.2$~pc
\citep{gusten&gwj_87, christopher&ssy_05} to $r_{\rm out} = 3.0$~pc.
These values are a compromise between the various estimates found 
in the literature: 
$0.8~{\rm pc} \leq r_{\rm in} \leq 1.4~{\rm pc}$ 
\citep{wright&cmh_01, genzel&ctw_85},
$2.5~{\rm pc} \leq r_{\rm out} \leq 2.7~{\rm pc}$ to the northeast 
\citep{wright&cmh_01, christopher&ssy_05, gusten&gwj_87}
and $4.2~{\rm pc} \leq r_{\rm out} \leq 7~{\rm pc}$ to the southwest 
\citep{gusten&gwj_87, serabyn&gww_86}.
We also assume that the CNR has a trapezoidal cross-section,
with axial thickness increasing from $h(r_{\rm in}) = 0.4$~pc 
to $h(r_{\rm out}) = 1.0$~pc \citep[in accordance with 
the inner value and the slope derived by][]{gusten&gwj_87}.
The volume of the CNR is then 
$V_{\rm \scriptscriptstyle CNR} = 18~{\rm pc}^3$.
Lastly, we take the plane of the CNR to lie at inclination $70^\circ$ 
to the plane of the sky \citep{genzel&ctw_85, jackson&ggr_93}
and to intersect the latter at position angle $25^\circ$
east of north \citep{jackson&ggr_93}, i.e., somewhat less 
than the position angle $31\fdg40$ of the Galactic plane.

\begin{figure}[!t]
\centering
\hspace*{-2cm}
\includegraphics[scale=0.85]{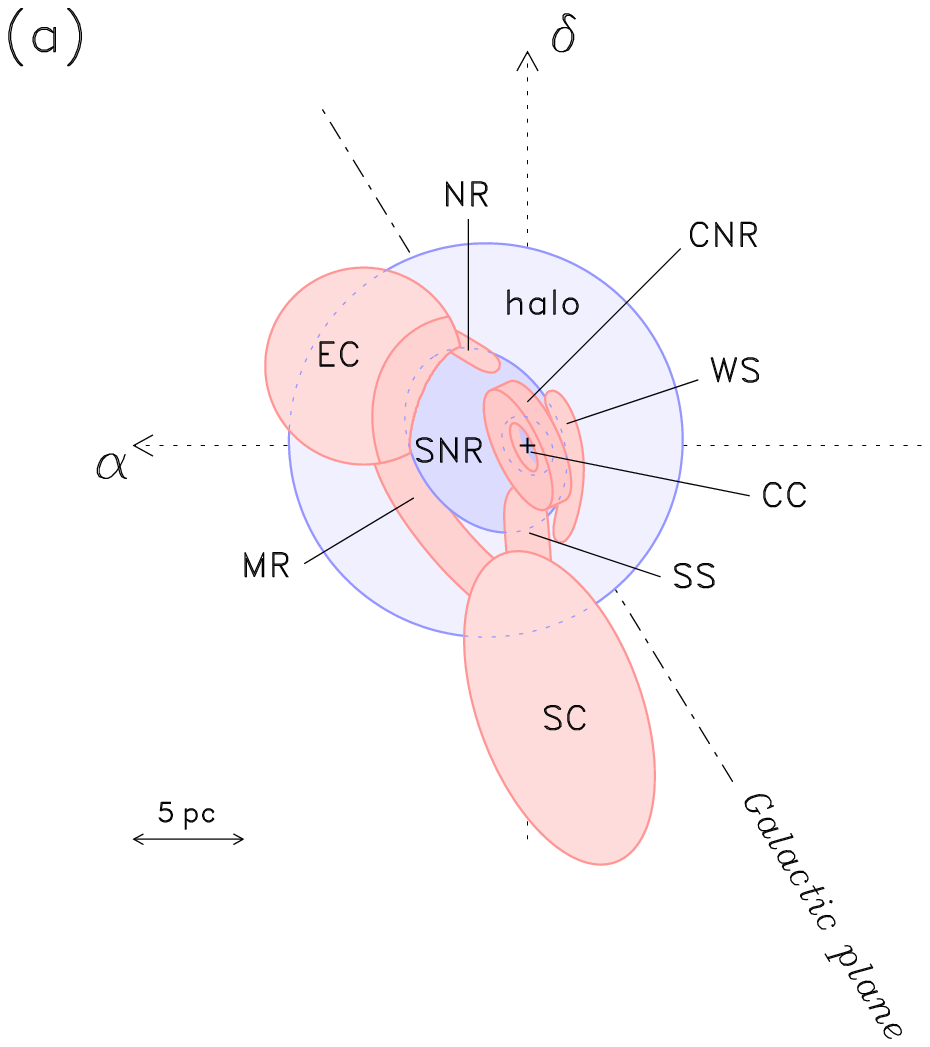}
\centering
\hspace*{-2cm}
\includegraphics[scale=0.85]{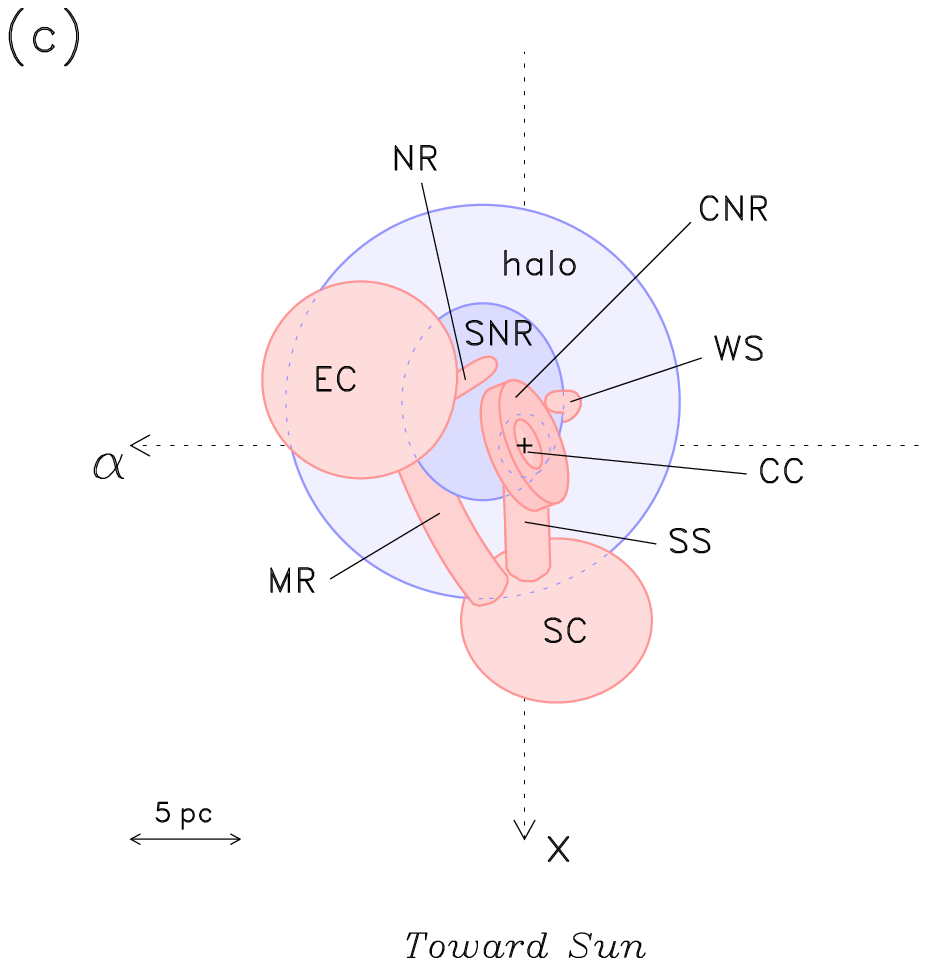}
\end{figure}
\begin{figure}[!t]
\centering
\hspace*{-2cm}
\includegraphics[scale=0.85]{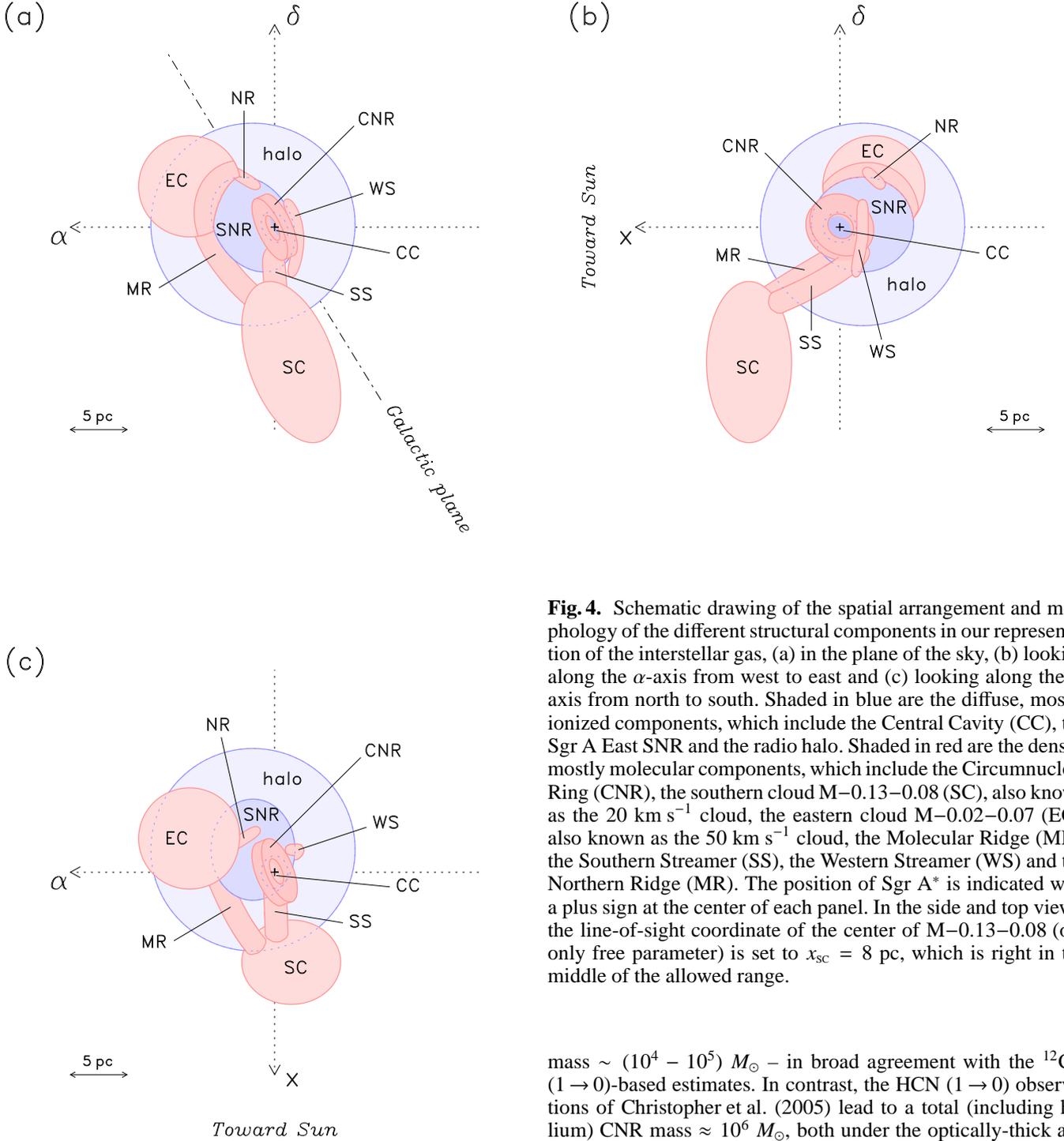}
\caption{
\label{geometry}
Schematic drawing of the spatial arrangement and morphology of 
the different structural components in our representation of 
the interstellar gas,
(a) in the plane of the sky, 
(b) looking along the $\alpha$-axis from west to east
and (c) looking along the $\delta$-axis from north to south.
Shaded in blue are the diffuse, mostly ionized components, which include
the Central Cavity (CC), the Sgr~A East SNR and the radio halo. 
Shaded in red are the denser, mostly molecular components, which include
the Circumnuclear Ring (CNR),
the southern cloud M$-$0.13$-$0.08 (SC), 
also known as the $20~{\rm km~s}^{-1}$ cloud,
the eastern cloud M$-$0.02$-$0.07 (EC),
also known as the $50~{\rm km~s}^{-1}$ cloud,
the Molecular Ridge (MR), the Southern Streamer (SS), 
the Western Streamer (WS) and the Northern Ridge (MR).
The position of Sgr~A$^*$ is indicated with a plus sign at the center
of each panel.
In the side and top views, the line-of-sight coordinate of the center
of M$-$0.13$-$0.08 (our only free parameter) is set to 
$x_{\rm \scriptscriptstyle SC} = 8$~pc,
which is right in the middle of the allowed range.
}
\end{figure}

It is interesting to note that the inner radius of the CNR 
($r_{\rm in} = 1.2$~pc) is somewhat smaller than the horizontal radius 
of the Central Cavity ($l_x / 2 = 1.45$~pc), 
so that the CNR slightly encroaches upon the Central Cavity.
On the other hand, the inner radius of the CNR coincides exactly with 
the semimajor axis of the nearly circular Western Arc 
\citep[$a = 1.2$~pc; see][]{zhao&mga_09}
and both structures have comparable inclinations ($70^\circ$ and
$63^\circ$, respectively), consistent with the Western Arc being 
the ionized inner edge of the western portion of the CNR. 

The molecular gas density inside the CNR varies over orders of magnitude, 
as revealed by the widely different values inferred from different tracers.
Typically, rotational lines of CO, CS, and HCN or HCO$^+$
yield true H$_2$ densities of a few $10^4~{\rm cm}^{-3}$
\citep{harris&jsg_85, sutton&djm_90, bradford&snb_05}, 
a few $10^5~{\rm cm}^{-3}$ \citep{serabyn&gww_86},
and a few $10^5~{\rm cm}^{-3}$ to a few $10^7~{\rm cm}^{-3}$
\citep{jackson&ggr_93, christopher&ssy_05, montero&hh_09}, respectively.
The molecular gas temperature also varies, although by a much smaller factor.
Measured values typically range from $\approx (50 - 200)$~K 
\citep[from HCN line ratios;][]{jackson&ggr_93}
through $\approx (100 - 200)$~K \citep[from CO line ratios;][]{sutton&djm_90}
and up to $\approx 300$~K \citep[from higher CO line ratios;][]{harris&jsg_85, 
bradford&snb_05}.
For convenience, we ignore all spatial variations (associated with
either clumping or large-scale gradients)
and we consider that the molecular gas has uniform density and temperature.
For the temperature, an obvious choice is the intermediate value 
$T_{\rm m} = 150$~K.
For the density, the assumption of uniformity means that 
true densities become irrelevant and that the appropriate quantity is
the space-averaged density, which we derive from the ratio of the CNR mass 
to its volume in the next paragraph.

The molecular mass of the CNR remains extremely uncertain.
$^{12}$CO $(1 \! \to \! 0)$ intensity measurements suggest 
that the CNR has an H$_2$ mass of a few $10^4~M_\odot$ 
\citep{genzel&ctw_85, serabyn&gww_86}.
The HCN $(3 \! \to \! 2)$ analysis of \cite{jackson&ggr_93} yields
a space-averaged H$_2$ density $\sim (10^4 - 10^5)~{\rm cm}^{-3}$,
which, multiplied by the volume 
$V_{\rm \scriptscriptstyle CNR} = 18~{\rm pc}^3$,
implies an H$_2$ mass $\sim (10^4 - 10^5)~M_\odot$ -- 
in broad agreement with the $^{12}$CO $(1 \! \to \! 0)$-based estimates.
In contrast, the HCN $(1 \! \to \! 0)$ observations of 
\cite{christopher&ssy_05} 
lead to a total (including helium) CNR mass $\approx 10^6~M_\odot$,
both under the optically-thick and virial assumptions,
and a similar virial mass $\approx 1.3 \times 10^6~M_\odot$ is obtained 
with the HCN $(4 \! \to \! 3)$ data of \cite{montero&hh_09}.
The discrepancy between \cite{christopher&ssy_05} 
(for their optically-thick mass) and \cite{jackson&ggr_93} 
can obviously be attributed to the adopted HCN-to-H$_2$ ratios
differing by a factor of 20,
whereas the discrepancy between \cite{christopher&ssy_05} 
(for their virial mass) and \cite{genzel&ctw_85, serabyn&gww_86}
can be explained if either the $^{12}$CO-based estimations miss 
a significant fraction of the mass due to gas concentration to very dense 
cores or if the dense HCN cores are out of virial equilibrium.
In the same fashion, \cite{oka&nkt_11} found a large discrepancy between 
the H$_2$ mass of the CNR inferred from the measured $^{13}$CO 
$(1 \! \to \! 0)$ intensity, $\approx (2.3-5.2) \times 10^5~M_\odot$, 
and its virial mass, $\approx 5.7 \times 10^6~M_\odot$,
which led them to reject the virial assumption.
Guided by this conclusion, we choose to disregard the virial estimates, 
and we adopt for the H$_2$ mass of the CNR
$(M_{\rm H_2})_{\rm \scriptscriptstyle CNR} = 2 \times 10^5~M_\odot$,
as a compromise between the $^{12}$CO $(1 \! \to \! 0)$- 
and $^{13}$CO $(1 \! \to \! 0)$-based estimates.
Upon dividing by $V_{\rm \scriptscriptstyle CNR} = 18~{\rm pc}^3$,
we then obtain for the space-averaged H$_2$ density in the CNR
$(\langle n_{\rm H_2} \rangle)_{\rm \scriptscriptstyle CNR} = 
2.2 \times 10^5~{\rm cm}^{-3}$.

The CNR also contains atomic gas, which tends to be confined 
to a photo-dissociated inner layer \citep{genzel&ctw_85, latvakoski&sgh_99}.
Here, we assume that the atomic layer extends radially over 0.4~pc
\citep{latvakoski&sgh_99}, i.e., from $r_{\rm in} = 1.2$~pc 
to $r_{\rm out,a} = 1.6$~pc, thereby occupying a volume of 1.65~pc$^3$.
We then adopt an H\,{\sc i} mass 
$(M_{\rm H{\scriptscriptstyle I}})_{\rm \scriptscriptstyle CNR} = 
1\,300~M_\odot$ \citep{latvakoski&sgh_99}, 
so that the space-averaged H\,{\sc i} density in the atomic layer is
$(\langle n_{\rm H{\scriptscriptstyle I}} \rangle)_{\rm \scriptscriptstyle CNR} 
= 3.2 \times 10^4~{\rm cm}^{-3}$, intermediate between the values 
$\approx 1.6 \times 10^4~{\rm cm}^{-3}$
and $\approx 4.0 \times 10^4~{\rm cm}^{-3}$
derived from the measured column densities 
toward the southwest and northeast ends 
of the photo-dissociated layer \citep{latvakoski&sgh_99}.
And for the temperature of the atomic gas, we take $T_{\rm a} = 300$~K 
\citep{genzel&ctw_85}.

\subsection{\label{model_shell}The Sgr~A East SNR}

In line with mainstream thought, we consider that Sgr~A East is an SNR,
with the observed radio synchrotron shell delimiting the swept-out cavity.
To determine the relevant geometrical parameters, 
we rely on the 20~cm continuum map of \cite{ekers&gsg_83},
which is expected to more correctly show the full extent of the radio shell
than the lower-frequency 90~cm continuum map of \cite{pedlar&aeg_89}.
If we assume that, similarly to the Central Cavity, the radio shell 
is ellipsoidal and axisymmetric about the vertical axis,
we may set its dimensions to $L_x \times L_y \times L_z 
= 9.0~{\rm pc} \times 9.0~{\rm pc} \times 6.7~{\rm pc}$.
Accordingly, the SNR cavity has a volume
$V_{\rm \scriptscriptstyle SNR} = 285~{\rm pc}^3$.

Unlike the Central Cavity and the CNR, the radio shell is not centered 
on Sgr~A$^*$, but on a slightly offset point 
$(x_{\rm c}, y_{\rm c}, z_{\rm c})$.
An eyeball location of the radio shell's projected center
in \citeauthor{ekers&gsg_83}'s (\citeyear{ekers&gsg_83}) 20~cm map
gives $y_{\rm c} = 1.2$~pc and $z_{\rm c} = -1.5$~pc, 
corresponding to a projected offset of 1.9~pc, 
which is somewhat less than the (rescaled) value $\approx 2.1$~pc 
quoted by the authors.
The determination of the line-of-sight offset, $x_{\rm c}$, 
is a little more tricky.
Here, we proceed from the premise \citep[see][]{yusef&mw_00,
maeda&bfm_02, herrnstein&h_05}
that the Central Cavity lies entirely inside the radio shell
and very close to its front surface.
The best value (rounded to 0.1~pc) leading to this particular
configuration is $x_{\rm c} = -2.0$~pc.

The SNR cavity contains hot ionized gas 
from both stellar ejecta and shocked interstellar matter.
The former is concentrated within a central $\approx (1.7-3.2)$~pc diameter 
core \citep{maeda&bfm_02, sakano&wdp_04, park&mbm_05}
and has a total mass $\approx (1.4-2)~M_\odot \, \phi_{\rm h}^{1/2}$
\citep{maeda&bfm_02, sakano&wdp_04},
which represents only small fractions of the total volume and mass 
of hot gas inside the entire SNR cavity. 
This entitles us to treat all the interior hot gas as if it were 
of interstellar origin and to assign it a temperature $T_{\rm h} = 1.3$~keV
($1.5 \times 10^7$~K) and solar abundances, as obtained for a "plume"
of shocked interstellar matter by \cite{park&mbm_05}.
However, we may not take up their derived density of the "plume",
which is almost certainly higher than average.
Instead, we assume that the interior hot gas has a total mass of 
$(27~M_\odot) \, \phi_{\rm h}^{1/2}$ \citep{koyama&uhm_07}
and that it completely fills ($\phi_{\rm h} = 1$) the volume 
of the SNR cavity outside the Central Cavity and the CNR, i.e., 
a volume of $260~{\rm pc}^3$.
The H$^+$ mass of hot gas inside the SNR cavity is then 
$(M_{\rm H^+})_{\rm hot} = 19~M_\odot$
and its (true or space-averaged) H$^+$ density 
$(n_{\rm H^+})_{\rm h} = 3.0~{\rm cm}^{-3}$.

\subsection{\label{model_halo}The radio halo}

The shape and size of the radio halo surrounding Sgr~A East 
are fairly well established observationally.
However, the fraction of the radio emission that can be attributed 
to thermal (warm ionized) gas is very uncertain, 
with some authors \citep[e.g.,][]{roy&r_09} going so far as 
to question the very need for a thermal contribution.
For our part, we regard the observational evidence for the presence 
of warm ionized gas as solid, and we assume that thermal and nonthermal 
gases are uniformly mixed throughout the radio halo. 
We model the latter as a sphere concentric with Sgr~A East 
\citep[see][]{yusef&m_87a}, which approximates
the $7'$ triangular halo of \cite{pedlar&aeg_89} best.
The result is a spherical halo centered on 
$(x_{\rm c}, y_{\rm c}, z_{\rm c})
= (-2.0~{\rm pc}, 1.2~{\rm pc}, -1.5~{\rm pc})$
and having diameter $d_{\rm halo} = 18$~pc
and volume $V_{\rm halo} = 3\,050~{\rm pc}^3$.
The volume available to warm ionized gas within this halo 
is reduced by the presence of the Sgr~A East SNR, the CNR 
and the other local molecular clouds (see Section~\ref{model_clouds})
to $V_{\rm wi} = 2\,440~{\rm pc}^3$.
With a projected surface area of $255~{\rm pc}^2$, the mean 
line-of-sight depth of warm ionized gas in the halo is then 9.5~pc.

For the temperature of the warm ionized gas, 
we choose $T_{\rm wi} = 7\,000$~K, 
which is intermediate between the temperatures used by 
\cite{pedlar&aeg_89} and \cite{anantharamaiah&pg_99}
and which is equal to the temperature adopted in Section~\ref{model_cavity} 
for the warm ionized gas inside the Central Cavity.
This choice of temperature requires an up-scaling of the emission measure
obtained by \cite{pedlar&aeg_89} to $4.3 \times 10^5~{\rm pc~cm}^{-6}$.
With a mean line-of-sight depth of 9.5~pc
and an assumed filling factor of unity,
the inferred (true or space-averaged) electron density is then
$(n_{\rm e})_{\rm halo} = 210~{\rm cm}^{-3}$.
If again all the free electrons come from hydrogen,
the (true or space-averaged) H$^+$ density in the halo is 
$(n_{\rm H^+})_{\rm halo} = (n_{\rm e})_{\rm halo}$
and the H$^+$ mass of the halo 
$(M_{\rm H^+})_{\rm halo} = 1.3 \times 10^4~M_\odot$.

Our electron density is somewhat higher than that derived
by \cite{pedlar&aeg_89}, because we adopted a higher temperature 
and hence found a larger emission measure.
However, our H$^+$ mass is considerably greater (by a factor $\simeq 6$).
Obviously, the mass difference is partly due to our higher density,
but the lion's share comes from our assumption that warm ionized gas
fills the entire 18~pc diameter radio halo (outside Sgr~A East 
and molecular clouds), as opposed to only its central 10~pc.
The above comparison underscores the important uncertainties 
in the actual volume and mass of the halo gas.

\subsection{\label{model_clouds}The belt of molecular clouds}

The nomenclature employed to distinguish different entities 
in the molecular belt around the Sgr~A complex is not unique.
The first two clouds unambiguously identified were 
M$-$0.02$-$0.07 to the east of Sgr~A$^*$ and M$-$0.13$-$0.08 to the south 
\citep{solomon&spw_72, guesten&wp_81}.
In NH$_3$ emission, M$-$0.02$-$0.07 peaks at 
$(l,b) \simeq (-0.02^\circ, -0.07^\circ)$, corresponding to
$(y,z) \simeq (5.5~{\rm pc}, -3.5~{\rm pc})$,
and M$-$0.13$-$0.08 peaks at
$(l,b) \simeq (-0.13^\circ, -0.08^\circ)$, corresponding to
$(y,z) \simeq (-11~{\rm pc}, -5~{\rm pc})$
\citep{guesten&wp_81}.
In CS emission, the dense core of M$-$0.02$-$0.07 peaks at
$(\Delta \alpha, \Delta \delta) \approx (3\fmn0,1\fmn5)$
with respect to Sgr~A$^*$, corresponding to
$(y,z) \approx (7~{\rm pc}, -4.5~{\rm pc})$ 
\citep{serabyn&la_92}.
The above transformations from $(l,b)$ to $(y,z)$
and from $(\Delta \alpha, \Delta \delta)$ to $(y,z)$
were made using the $(l_{\rm A^*}, b_{\rm A^*}) = 
(-0^\circ 03' 20.\!''\!5, -0^\circ 02' 46.\!''\!3)$
coordinates of Sgr~A$^*$ and the $58\fdg60$ angle 
between the $(\alpha,\delta)$ and $(y,z)$ systems
(see Section~\ref{intro}).

The line-of-sight positions of both clouds are still under debate.
The general belief is that M$-$0.13$-$0.08 lies in front of Sgr~A$^*$ 
\citep{zylka&mw_90, park&mbm_04}, the CNR \citep{coil&h_99} 
and Sgr~A East \citep{herrnstein&h_05}.
Besides, M$-$0.13$-$0.08 was argued to be less than $\approx 9$~pc 
away from Sgr~A East \citep{coil&h_00}.
Regarding M$-$0.02$-$0.07, a large portion of the cloud appears 
to lie behind Sgr~A East, but there is also evidence that the cloud 
extends all the way to the near side of Sgr~A East
\citep{serabyn&la_92, coil&h_00, herrnstein&h_05, lee&pcd_08}.

Curving around the eastern edge of Sgr~A East is a prominent molecular 
feature, usually referred to as the Molecular Ridge (or Curved Streamer).
Some authors consider that the Molecular Ridge actually belongs to 
M$-$0.02$-$0.07 and coincides with the fraction of the cloud 
that has been swept up and compressed by the Sgr~A East forward shock
\citep[e.g.,][]{serabyn&la_92, maeda&bfm_02}.
In this scenario, the Molecular Ridge must adhere to Sgr~A East 
and form the eastern part of the gas-and-dust shell surrounding it.
Other authors regard the Molecular Ridge as a separate cloud
that connects M$-$0.02$-$0.07 to M$-$0.13$-$0.08,
independent of Sgr~A East
(e.g., \citeauthor{herrnstein&h_05} \citeyear{herrnstein&h_05};
\citeauthor{lee&pcd_08} \citeyear{lee&pcd_08};
see also \citeauthor{zylka&mw_90} \citeyear{zylka&mw_90},
who argue that the Molecular Ridge (their Curved Streamer) 
probably makes a true connection only with M$-$0.13$-$0.08).
An intermediate possibility is that the Molecular Ridge is divided
into a northern half that represents the shock-compressed portion 
of M$-$0.02$-$0.07 and a southern half that splits off the edge 
of Sgr~A East and continues south toward M$-$0.13$-$0.08 \citep{coil&h_00}.

Since Sgr~A East is, by all accounts, impacting upon M$-$0.02$-$0.07,
a fraction of the cloud must necessarily be shock-compressed 
into a piece of shell, which in turn must show up as a ridge
around the eastern edge of Sgr~A East.
It is, therefore, hard to escape the conclusion that the observed 
Molecular Ridge comprises shock-compressed material from M$-$0.02$-$0.07.
On the other hand, the Molecular Ridge appears to extend 
toward M$-$0.13$-$0.08 past the boundary of M$-$0.02$-$0.07,
so that it must also contain material that is {\it not} from M$-$0.02$-$0.07.
This material could either be another piece of the shock-compressed shell 
surrounding Sgr~A East or form a connecting bridge to M$-$0.13$-$0.08.  
Based on existing evidence for a physical connection between 
M$-$0.02$-$0.07 and M$-$0.13$-$0.08 \citep{lee&pcd_08}, 
we give preference to the second possibility.

Three other elongated molecular features exist in the region of interest.
Due south of Sgr~A$^*$, the Southern Streamer stretches between the CNR 
and M$-$0.13$-$0.08 and appears to link them together \citep{coil&h_99}.
West and northeast of Sgr~A$^*$, the Western Streamer and 
the Northern Ridge follow (closely in the case of the Western Streamer) 
the contour of Sgr~A East; 
both could be pieces of the shock-compressed shell surrounding Sgr~A East
\citep{mcgary&co_01}.

The projected dimensions of the above molecular clouds
have been given a range of values.
\cite{solomon&spw_72} estimated the diameters of M$-$0.02$-$0.07 
and M$-$0.13$-$0.08 at $\sim (15-50)$~pc.
\cite{zylka&mw_90} found that  M$-$0.13$-$0.08 
is $\approx 15~{\rm pc} \times 7.5~{\rm pc}$ in size
and the Molecular Ridge (their Curved Streamer) $\approx 7.5$~pc wide in $b$.
According to \cite{coil&h_00, coil&h_99}, the Molecular Ridge 
and the Southern Streamer are both $\approx 2$~pc wide
and $\gtrsim 12~{\rm pc}$ and $\approx 10$~pc long in projection, 
respectively,
while according to \cite{mcgary&co_01}, the Western Streamer 
has a north-south extent $\approx 7$~pc and the Northern Ridge 
a northeast-southwest extent $\approx 3.5$~pc.
No direct information is available on the line-of-sight dimensions
of these clouds.

Based on the above elements and on existing maps of the GC region,
we represent the morphology and layout of the main molecular clouds 
in the following way:

M$-$0.13$-$0.08 is approximated as a $15~{\rm pc} \times (7.5~{\rm pc})^2$ 
ellipsoid \citep[see][]{zylka&mw_90}, with long axis in the plane of the sky, 
at position angle $20^\circ$ east of north, i.e., roughly parallel 
to the trace of the Galactic plane.
Its volume is $V_{\rm \scriptscriptstyle SC} = 442~{\rm pc}^3$,
where subscript SC stands for southern cloud.
In projection, the cloud is located south of Sgr~A$^*$, 
just below the southern boundary of Sgr~A East, 
and its center, identified with the NH$_3$ emission peak, is at
$(y_{\rm \scriptscriptstyle SC}, z_{\rm \scriptscriptstyle SC}) 
= (-11~{\rm pc}, -5~{\rm pc})$ \citep{guesten&wp_81}.
Along the line of sight, the cloud lies completely in front of Sgr~A$^*$,
within a three-dimensional distance of Sgr~A East $\lesssim 9$~pc 
\citep{coil&h_00}.
This double constraint restricts the line-of-sight coordinate 
of the cloud center to the fairly loose range (in round numbers)
$4~{\rm pc} \le x_{\rm \scriptscriptstyle SC} \le 12~{\rm pc}$.

M$-$0.02$-$0.07 is most easily described starting from its original,
pre-explosion shape, which we approximate as a 9~pc diameter sphere.
In projection, this sphere is located at the eastern boundary of Sgr~A East;
its center is not identified with the NH$_3$ emission peak, 
which refers to the full present-day M$-$0.02$-$0.07 cloud 
(including its shock-compressed portion),
but rather with the CS emission peak, 
which pertains to the preshock core alone and is at 
$(y_{\rm \scriptscriptstyle EC}, z_{\rm \scriptscriptstyle EC}) 
= (7~{\rm pc}, -4.5~{\rm pc})$ \citep{serabyn&la_92}.
Along the line of sight, the sphere extends from the near side 
to the far side of the eastern part of Sgr~A East,
with a small displacement toward the back; 
we place its center 1~pc behind the center of Sgr~A East,
at $x_{\rm \scriptscriptstyle EC} = -3~{\rm pc}$.
Here, subscript EC stands for eastern cloud.

Obviously, the fraction of the 9~pc diameter sphere that overlaps 
with Sgr~A East has been cleared of gas by the supernova explosion, 
such that the swept-up gas now resides in a piece of shell 
squeezed between the preshock core and the Sgr~A East cavity.
The present-day M$-$0.02$-$0.07 cloud is thus composed of 
the preshock core and the piece of swept-up shell.
It occupies the volume of the 9~pc diameter sphere outside Sgr~A East, 
$V_{\rm \scriptscriptstyle EC} = 356~{\rm pc}^3$, 
which is split between $V_{\rm {\scriptscriptstyle EC},shell} 
= 51~{\rm pc}^3$ for the piece of shell,
assumed to have 1.5 times the preshock density (see below),
and $V_{\rm {\scriptscriptstyle EC},core} = 305~{\rm pc}^3$ 
for the preshock core.
With the above volume, the piece of shell must be 1.5~pc thick,
consistent with existing maps of M$-$0.02$-$0.07.

The Molecular Ridge consists of the swept-up shell from M$-$0.02$-$0.07 
and a connecting bridge to M$-$0.13$-$0.08.
The "Bridge" is modeled as a curved cylinder, with diameter 2~pc
\citep{coil&h_00}, extending from the southern part of M$-$0.02$-$0.07 
to the northeastern end of M$-$0.13$-$0.08 along the southeastern edge 
of Sgr~A East.
The total length of the Bridge depends on the line-of-sight position 
of M$-$0.13$-$0.08, being between
9~pc (for $x_{\rm \scriptscriptstyle SC} = 4~{\rm pc}$)
and 15~pc (for $x_{\rm \scriptscriptstyle SC} = 12~{\rm pc}$).
Correspondingly, the volume of the Bridge is between 
$V_{\rm bridge} = 28~{\rm pc}^3$ and $47~{\rm pc}^3$.

The three other streamers are also modeled as curved cylinders,
with diameters 2~pc for the Southern Streamer \citep{coil&h_99}
and 1~pc for the Western Streamer and the Northern Ridge
\citep[as estimated from the NH$_3$ maps of][]{mcgary&co_01}.

\noindent
$\bullet$ The Southern Streamer connects the northern end 
of M$-$0.13$-$0.08 to the southeastern part of the CNR.
Its total length, which again depends on the line-of-sight position 
of M$-$0.13$-$0.08, is between
7~pc (for $x_{\rm \scriptscriptstyle SC} = 4~{\rm pc}$)
and 14~pc (for $x_{\rm \scriptscriptstyle SC} = 12~{\rm pc}$),
and its volume between $V_{\rm \scriptscriptstyle SS} = 22~{\rm pc}^3$ 
and $44~{\rm pc}^3$.

\noindent
$\bullet$ The Western Streamer follows the western surface of Sgr~A East,
at the line-of-sight distance of Sgr~A East's center, $x_{\rm c} = -2$~pc
(see footnote~\ref{foot}). 
Its total length is 8~pc 
and its volume $V_{\rm \scriptscriptstyle WS} = 6~{\rm pc}^3$.

\noindent
$\bullet$ The Northern Ridge extends along the northern surface of Sgr~A East,
in the northeast-southwest direction.
Along the line-of-sight, it cannot possibly be located in front of 
the northeastern part of the CNR \citep[as suggested by][]{mcgary&co_01};
instead it lies close to \citep{herrnstein&h_05},
but behind \citep{lee&pcd_08} the center of Sgr~A East.
Its total length is 4~pc and its volume 
$V_{\rm \scriptscriptstyle NR} = 3~{\rm pc}^3$.

We now turn to the physical conditions in the belt of molecular clouds.
The gas temperature has mostly been inferred from NH$_3$ line ratios,
leading to $\approx (50 - 120)$~K in M$-$0.02$-$0.07 and M$-$0.13$-$0.08
\citep{guesten&wp_81},
$\approx (17 - 35)$~K with a localized jump to $\approx 300$~K 
in the Southern Streamer \citep{coil&h_99},
and $\sim 15$~K for $\sim 75\%$ of the gas {\it versus} 
$\sim 200$~K for the other $\sim 25\%$ across the central 10~pc
\citep{herrnstein&h_05}.
Like for the other objects, we neglect the important temperature
variations and adopt a uniform temperature, to which we naturally
assign the mean value of \cite{herrnstein&h_05}, $T_{\rm m} = 60$~K.
This value is within the ranges derived by \cite{guesten&wp_81}
and \cite{coil&h_99}.

Little is known on either the true or space-averaged gas density 
in the molecular belt.
From their CS observations of M$-$0.02$-$0.07, \cite{serabyn&la_92} 
estimated the true H$_2$ density in the cloud at 
$\approx (1-2) \times 10^6~{\rm cm}^{-3}$,
and with an assumed line-of-sight depth $\approx 2.5$~pc,
they obtained a space-averaged H$_2$ density
$\approx 1.5 \times 10^4~{\rm cm}^{-3}$ in the Molecular Ridge
and $\approx 3 \times 10^4~{\rm cm}^{-3}$ near the peak 
of the preshock core.
Although a 2.5~pc line-of-sight depth is probably reasonable for 
the Molecular Ridge, a more appropriate choice for the peak region
would be the 9~pc diameter of the preshock core, which would lower 
the space-averaged H$_2$ density to $\approx 10^4~{\rm cm}^{-3}$.
For comparison, NH$_3$ observations of the Southern Streamer 
and the Molecular Ridge by \cite{coil&h_99,coil&h_00} led to 
a space-averaged H$_2$ density $\sim (1-2) \times 10^5~{\rm cm}^{-3}$
in both streamers, assuming a line-of-sight depth $\approx 2$~pc
and an NH$_3$-to-H$_2$ ratio of $10^{-8}$.
With a presumably more realistic NH$_3$-to-H$_2$ ratio of $10^{-7}$
\citep{herrnstein&h_05}, the space-averaged H$_2$ density would be 
$\sim (1-2) \times 10^4~{\rm cm}^{-3}$,
in very good agreement with \cite{serabyn&la_92}.
This good agreement prompts us to adopt 
$(\langle n_{\rm H_2} \rangle)_{\rm streamer} = 1.5 \times 10^4~{\rm cm}^{-3}$
in the four streamers, namely, the Molecular Ridge, the Southern Streamer,
the Western Streamer and the Northern Ridge.
For the preshock core of M$-$0.02$-$0.07 and for M$-$0.13$-$0.08,
we rely on our downward revision of the space-averaged H$_2$ density 
near the peak of the preshock core to adopt
$(\langle n_{\rm H_2} \rangle)_{\rm {\scriptscriptstyle EC},core} 
= (\langle n_{\rm H_2} \rangle)_{\rm \scriptscriptstyle SC}
= 10^4~{\rm cm}^{-3}$.
It is certainly reassuring to have a space-averaged density 
lower in the preshock core of M$-$0.02$-$0.07 than in the Molecular Ridge 
(which includes the postshock portion of M$-$0.02$-$0.07).

Combining our adopted space-averaged densities with the volumes derived
above, we obtain the H$_2$ masses listed in Table~\ref{summary_thermo}.
Our hydrogen masses of M$-$0.13$-$0.08 and M$-$0.02$-$0.07
(either the entire cloud or its preshock core alone)
are in broad agreement with existing estimates, 
which include $\gtrsim 10^5~M_\odot$ for both clouds
\citep[from a CO emission map;][]{solomon&spw_72},
$\approx 3 \times 10^5~M_\odot$ for M$-$0.13$-$0.08 
and $\gtrsim 2 \times 10^5~M_\odot$ for the Sgr A East Core
\citep[from dust emission maps;][]{zylka&mw_90},
$\approx 1.5 \times 10^5~M_\odot$ for M$-$0.02$-$0.07
\citep[from CS emission maps;][]{serabyn&la_92}
and $\sim 2 \times 10^5~M_\odot$ for the core of M$-$0.02$-$0.07
\citep[from NH$_3$ emission maps;][]{herrnstein&h_05}.
Similarly, our derived hydrogen mass of the Molecular Ridge lies well within 
the existing range, which includes
$\approx (1-1.5) \times 10^5~M_\odot$ 
\citep[from dust emission maps;][]{zylka&mw_90},
$\sim 1.5 \times 10^4~M_\odot$
\citep[from NH$_3$ emission maps, after rescaling to 
a NH$_3$-to-H$_2$ ratio of $10^{-7}$;][]{coil&h_00}
and $\gtrsim 3 \times 10^4~M_\odot$ 
\citep[also from NH$_3$ emission maps;][]{herrnstein&h_05}.
In contrast, our derived hydrogen mass of the Southern Streamer 
is somewhat lower than the two existing NH$_3$ estimates, 
$\sim 3.5 \times 10^4~M_\odot$ \citep[after rescaling to 
a NH$_3$-to-H$_2$ ratio of $10^{-7}$;][]{coil&h_99}
and $\gtrsim 8 \times 10^4~M_\odot$ \citep{herrnstein&h_05}.
This slight discrepancy finds its origin in the very definition 
of the Southern Streamer.
For instance, \cite{coil&h_99} considered a longer structure than we did, 
which extends deeper into M$-$0.13$-$0.08;
what they regarded as the southern part of the Southern Streamer 
actually belongs to our M$-$0.13$-$0.08 cloud.
Finally, for the Western Streamer and the Northern Ridge,
we find hydrogen masses that are very close to the NH$_3$ estimates,
$\sim 4 \times 10^3~M_\odot$ and $\sim 2 \times 10^3~M_\odot$,
respectively \citep{herrnstein&h_05}.

\section{\label{conclu}Conclusions}

Motivated by the recognized impact of the massive black hole
at the dynamical center of our Galaxy and by the crucial role 
played by its interstellar environment,
and encouraged by the multitude of recent observational findings
at both long (radio and infrared) and short (X-ray and $\gamma$-ray)
wavelengths, we have tried to put some order in the current muddled view
of this inherently complex Galactic region.
We restricted our attention to the interstellar gas within $\sim 10$~pc 
of the central black hole, and we described it in terms of five distinct 
structural components:
the Central Cavity, the Circumnuclear Ring (CNR) encircling it, 
the Sgr~A East SNR encompassing both, the surrounding radio halo 
and the belt of massive molecular clouds stretching along the Galactic plane.
We first reviewed the existing observations of these five gaseous components.
We then integrated them as well as possible into a detailed three-dimensional 
representation of the interstellar gas, in which 
each component is assigned both geometrical (position, shape, dimensions)
and thermodynamic (phase, temperature, density, mass) characteristics.
These characteristics are summarized in Tables~\ref{summary_geo} 
and \ref{summary_thermo}, while the overall spatial disposition of 
the different components is graphically shown in Figure~\ref{geometry} 
from three orthogonal viewpoints.

In the process of constructing our gas representation, 
we inevitably came across conflicting observational claims. 
We naturally chose those that we found more convincing,
sometimes with the benefit of hindsight,
and/or those that appeared to fit the general picture better,
and we disregarded the others.
A synthetic list of important observational claims that we were led
to push aside in favor of stronger ones is given in Table~\ref{list}.

\begin{table*}[t]
\caption{Observational claims disregarded in our gas representation.
}
\label{list}
\begin{minipage}[t]{2\columnwidth}
\centering 
\renewcommand{\footnoterule}{}  
\begin{tabular}{lll}
\hline
\hline
\noalign{\smallskip}
& Claims & References \\
\noalign{\smallskip}
\hline
\noalign{\smallskip}
Central Cavity & 
\parbox[t]{11cm}{
The diffuse cm radio continuum emission from Sgr~A West is mainly nonthermal.
} &
\parbox[t]{4cm}{\cite{ekers&gsg_83}} \\
\noalign{\smallskip}
& \parbox[t]{11.3cm}{
The Western Arc and the Northern Arm are contained in a single 
one-armed linear spiral.
} &
\parbox[t]{4cm}{\cite{lo&c_83}} \\
\noalign{\smallskip}
& \parbox[t]{11cm}{
Ionized gas in the Northern Arm is on an unbound orbit about Sgr~A$^*$.
} &
\parbox[t]{4cm}{\cite{roberts&yg_96}\\
\cite{yusef&rb_98}\\ 
\cite{muzik&esm_07}} \\
\noalign{\smallskip}
& \parbox[t]{11cm}{
The Northern and Eastern Arms are the ionized outer rims 
of the Northern Streamer.
} &
\parbox[t]{4cm}{\cite{davidson&wwl_92}\\
\cite{jackson&ggr_93}} \\
\noalign{\smallskip}
\hline
\noalign{\smallskip}
CNR & 
\parbox[t]{11cm}{The CNR is off-centered from Sgr~A$^*$ 
and noticeably warped.
} &
\parbox[t]{4cm}{\cite{gusten&gwj_87}} \\
\noalign{\smallskip}
& \parbox[t]{11cm}{
Dense molecular cores inside the CNR are in virial equilibrium.
} &
\parbox[t]{4cm}{\cite{christopher&ssy_05}\\
\cite{montero&hh_09}} \\
\noalign{\smallskip}
\hline
\noalign{\smallskip}
Sgr~A East &
\parbox[t]{11cm}{
Sgr~A East is separated from Sgr~A West by a finite distance
along the line of sight.
} &
\parbox[t]{4.5cm}{\cite{karlsson&ssw_03}\\
\cite{sjouwerman&p_08}} \\
\noalign{\smallskip}
\hline
\noalign{\smallskip}
Radio halo & \parbox[t]{11cm}{
The radio halo is purely nonthermal.
} &
\parbox[t]{4cm}{\cite{roy&r_09}} \\
\noalign{\smallskip}
& \parbox[t]{11cm}{
The radio halo is located in front of Sgr~A East.
} &
\parbox[t]{4cm}{\cite{pedlar&aeg_89}} \\
\noalign{\smallskip}
\hline
\noalign{\smallskip}
Molecular belt &
\parbox[t]{11cm}{
The Curved Streamer (or Molecular Ridge) is a northward extension
of M$-$0.13$-$0.08, \\ separate from M$-$0.02$-$0.07.
} &
\parbox[t]{4cm}{\cite{zylka&mw_90}} \\
\noalign{\smallskip}
& \parbox[t]{11cm}{
The Molecular Ridge as a whole is the shock-compressed portion 
of M$-$0.02$-$0.07.
} &
\parbox[t]{4cm}{\cite{serabyn&la_92}\\
\cite{maeda&bfm_02}} \\
\noalign{\smallskip}
& \parbox[t]{11cm}{
The Molecular Ridge is a separate cloud
connecting M$-$0.02$-$0.07 to M$-$0.13$-$0.08.
} &
\parbox[t]{4cm}{\cite{herrnstein&h_05}\\
\cite{lee&pcd_08}} \\
\noalign{\smallskip}
& \parbox[t]{11cm}{
The Northern Ridge sits in front of the northeastern lobe of the CNR.
} &
\parbox[t]{4cm}{\cite{mcgary&co_01}} \\
\noalign{\smallskip}
& \parbox[t]{11cm}{
The Western Streamer is highly inclined to the plane of the sky.
} &
\parbox[t]{4cm}{\cite{herrnstein&h_05}} \\
\noalign{\smallskip}
& \parbox[t]{11cm}{
The $20~{\rm km~s}^{-1}$ cloud lies mostly behind Sgr~A West.
} &
\parbox[t]{4.5cm}{\cite{karlsson&ssw_03}\\
\cite{sjouwerman&p_08}} \\
\noalign{\smallskip}
\hline
\end{tabular}
\end{minipage}
\end{table*}

In our gas representation, the total interstellar hydrogen mass 
of the studied region amounts to $\approx 7 \times 10^5~M_\odot$.
As expected, most of this mass resides in the dense, molecular components,
with nearly 70\% in the molecular belt and nearly 30\% in the CNR.
The radio halo encloses $\approx 2\%$ of the mass,
the Central Cavity $\approx 0.05\%$
and the Sgr~A East SNR a negligible $\approx 0.003\%$.
Although the Central Cavity is commonly regarded as a spiral-shaped 
H\,{\sc ii} region, only half its mass is actually in the form
of warm ionized gas, and barely $\approx 3\%$ is truly contained 
in the Minispiral;
the other half of the mass belongs to neutral atomic streamers.

Our gas representation is merely meant to provide a three-dimensional 
snapshot of the innermost interstellar region as it presently stands,
not to explain how this highly interacting system actually works.
However, to set the general framework, we were led to briefly
discuss the physical interactions between the different gaseous components.
This also helped us to constrain their positional relationships
(primarily along the line of sight), 
in the light of the available observational data on their morphology,
kinematics, absorption-versus-emission properties, etc.

As we learned along the way, much of the present state of the innermost 
interstellar region is likely the direct result of two antagonistic 
phenomena, namely, unsteady accretion onto the central black hole
and rapid expansion triggered by a nearby supernova explosion.
The CNR is almost certainly a manifestation of the accretion,
and so are also several of the local streamers, including
the ionized Northern and Eastern Arms of the Minispiral,
their neutral counterparts in the Central Cavity
and, in a less obvious manner, the Southern Streamer, which was found 
to carry material from the M$-$0.13$-$0.08 GMC to the CNR.
Meanwhile, the supernova explosion may be held responsible 
for the existence of Sgr~A East and of most of the dense and elongated 
features observed along its projected boundary -- 
the gas-and-dust shell, all or part of the Molecular Ridge, 
the Western Streamer and probably the Northern Ridge.
Interestingly, the CNR itself might be a by-product of the explosion,
formed as the gas-and-dust shell swept up by the supernova shock 
passed over Sgr~A$^*$ and left a piece in its deep gravitational potential.
In this case, the CNR would offer a nice illustration of the complex
interplay between supernova-driven expansion and black-hole accretion.

A self-consistent and reliable three-dimensional snapshot 
of the innermost interstellar gas, which incorporates all relevant
information from diverse observational sources,
may serve several purposes.
First and foremost, such a snapshot may constitute an objective reference 
against which to test different scenarios of the intricate dynamics
of the region as well as of the detailed operation of the whole nuclear engine, 
with the precise role played by each of its (stellar and interstellar) 
constituents.
This, in turn, may pave the way to improve our understanding of
galactic nuclei in general.
In a different perspective, our gas representation may provide 
the necessary framework to study specific problems related to 
Sgr~A$^*$ and its immediate environment. 
As mentioned in the introduction, the problem that we personally 
have in mind bears on the propagation and annihilation of positrons 
from Sgr~A$^*$.

In its present form, our gas representation still suffers from many 
uncertainties.
Perhaps most uncertain of all is the line-of-sight location 
of the M$-$0.13$-$0.08 GMC, which we left as a free parameter 
(albeit restricted to a narrowed-down range).
Another uncertain aspect concerns the true nature of what we referred to 
as the Molecular Ridge -- whether all of it represents a piece of 
the shock-compressed shell surrounding Sgr~A East or whether 
its southern half forms a separate bridge between M$-$0.02$-$0.07 
and M$-$0.13$-$0.08.
Equally uncertain are the number of neutral streamers inside the Central Cavity
and their exact relationship with the ionized Northern and Eastern Arms 
of the Minispiral.
And of course, none of the thermodynamic parameters is really well constrained.
Hopefully, future observations will help to sort out the remaining 
uncertainties.

\begin{acknowledgement}{}
The author expresses her gratitude to N.~Guessoum, P.~Jean,
J.~ Kn\"odlseder, S.~Park and F.~Yusef-Zadeh
for enlightening discussions and help with the preparation of the manuscript.
\end{acknowledgement}

\bibliographystyle{aa}
\bibliography{BibTex}

\end{document}